\font \fiverm=cmr5
\font \sixrm=cmr6
\font \sevenrm=cmr7
\font \eightrm=cmr8

\font \tengoth=eufm10
\font \sevengoth=eufm7
\font \fivegoth=eufm5

\newfam\gothfam
\textfont \gothfam=\tengoth
\scriptfont \gothfam=\sevengoth
\scriptscriptfont \gothfam=\fivegoth

\newfam\srmfam
\textfont \srmfam=\eightrm
\scriptfont \srmfam=\sixrm
\scriptscriptfont \srmfam=\fiverm

\font \tengothb=eufb10
\font \sevengothb=eufb7
\font \fivegothb=eufb5

\newfam\gothbfam
\textfont \gothbfam=\tengothb
\scriptfont \gothbfam=\sevengothb
\scriptscriptfont \gothbfam=\fivegothb

\font \tenmath=msbm10
\font \sevenmath=msbm7
\font \fivemath=msbm5

\newfam\mathfam
\textfont \mathfam=\tenmath
\scriptfont \mathfam=\sevenmath
\scriptscriptfont \mathfam=\fivemath

\def\corde#1#2-#3{{#1}_{#2},\ldots ,{#1}_{#3}}

\def\ordcorde#1#2-#3{{#1}_{#2} \le \cdots \le {#1}_{#3}}
\def\strictordcorde#1#2-#3{{#1}_{#2} < \cdots < {#1}_{#3}}
\def \restr#1{\mathstrut_{\textstyle |}\raise-6pt\hbox{$\scriptstyle #1$}}
\def \srestr#1{\mathstrut_{\scriptstyle |}\hbox to -1.5pt{}\raise-4pt\hbox{$\scriptscriptstyle #1$}}

\def\dbar{d\!\!\hbox to 4.5pt{\hfill\vrule height 5.5pt depth -5.3pt
        width 3.5pt}}

\def\fleche#1{\mathop{\hbox to #1 mm{\rightarrowfill}}\limits}
\def\gfleche#1{\mathop{\hbox to #1 mm{\leftarrowfill}}\limits}
\def\inj#1{\mathop{\hbox to #1 mm{$\lhook\joinrel$\rightarrowfill}}\limits}
\def\ginj#1{\mathop{\hbox to #1 mm{\leftarrowfill$\joinrel\rhook$}}\limits}
\def\surj#1{\mathop{\hbox to #1 mm{\rightarrowfill\hskip 2pt\llap{$\rightarrow$}}}\limits}
\def\gsurj#1{\mathop{\hbox to #1 mm{\rlap{$\leftarrow$}\hskip 2pt \leftarrowfill}}\limits}
\def\semi{\mathrel{\times}\kern -1pt\joinrel\mathrel{\raise 1.2pt\hbox{${\scriptscriptstyle |}$}}}
\def \mop#1{\mathop{\hbox{\rm #1}}\nolimits}
\def \smop#1{\mathop{\hbox{\sevenrm #1}}\nolimits}

\def \mopl#1{\mathop{\hbox{\rm #1}}\limits}

\documentclass[10pt]{amsart}

\usepackage{amssymb}
\usepackage{amsthm}
\newtheorem{theorem}{Theorem}
\newtheorem{definition}{Definition}
\newtheorem{corollary}{Corollary}
\newtheorem{lemma}{Lemma}

\newtheorem{proposition}{Proposition}

\newtheorem{remark}{Remark}

\usepackage{enumerate}
\usepackage{amsfonts}
\usepackage{amsmath}

\def\un{\hbox{\bf 1}}

\begin{document}
%%%%%%%%%%%%%%%%%%%%%%%%%%%%%%%%%%%%%%%%%%%%%
%%%%%%%%%%%%%%%%%%%%%%%%%%%%%%%%%%%%%%%%%%%%%

\title[The combinatorics of Bogoliubov's recursion in renormalization]
      {The combinatorics of Bogoliubov's recursion in renormalization}

\author{Kurusch Ebrahimi-Fard}
\address{Laboratoire MIA, 
         Universit\'e de Haute Alsace, 
         4-6 rue des Fr\`eres Lumi\`ere, 
         68093 Mulhouse, C\'edex 07, France}
\email{kurusch.ebrahimi-fard@uha.fr}
\urladdr{http://www.th.physik.uni-bonn.de/th/People/fard/}

\author{Dominique Manchon}
\address{Universit\'e Blaise Pascal,
         C.N.R.S.-UMR 6620,
         63177 Aubi\`ere, France}
         \email{manchon@math.univ-bpclermont.fr}
         \urladdr{http://math.univ-bpclermont.fr/~manchon/}

%%%%%%%%%%%%%%%%%%%%%%%%%%%%%%%%%%%%%%%%%%%%%%%%%%%%%%%%%%%%%%%%%%%
\date{December 11, 2008\\
\noindent {\footnotesize{${}\phantom{a}$ 2001 PACS Classification:
03.70.+k, 11.10.Gh, 02.10.Hh}} }
%%%%%%%%%%%%%%%%%%%%%%%%%%%%%%%%%%%%%%%%%%%%%%%%%%%%%%%%%%%%%%%%%%%

\maketitle
%%%%%%%%%%%%%%%%%%%%%%%%%%%%%%%%%%%%%%%%%%%%%%%%%%%%%%%%%%%%%%%%%%%

\begin{abstract}
We describe various combinatorial aspects of the Birkhoff--Connes--Kreimer factorization in perturbative renormalisation. The analog of Bogoliubov's preparation map on the Lie algebra of Feynman graphs is identified with the pre-Lie Magnus expansion. Our results apply to any connected filtered Hopf algebra, based on the pro-nilpotency of the Lie algebra of infinitesimal characters.
\end{abstract}

\tableofcontents

%%%%%%%%%%%%%%%%%%%%%%%%%%%%%%%%%%%%%%%%%%%%%%%%%%%%%%%%%%%%%%%%%%%

\section{Introduction}
\label{sect:Intro}

The recursive nature of the process of renormalization in perturbative quantum field theory (with respect to the loop number of Feynman diagrams considered) has been settled long ago by N.~N.~Bogoliubov and O.~S.~Parasiuk \cite{BP,H,Z}. However, its mathematical structure has become much more transparent since the seminal discovery of a Hopf algebra structure on Feynman diagrams by D.~Kreimer \cite{kreimer1998} and subsequent work by A.~Connes and Kreimer~\cite{CK,CK1,CK2}.

\smallskip

We present here a concise survey of several recent works on algebro-combinatorial aspects of the process of perturbative renormalization, in particular Bogoliubov's recursion respectively Connes--Kreimer's Birkhoff decomposition. The natural mathematical setting for such studies is provided by connected filtered Hopf algebras. Indeed, this leads, with the connectedness property being the very key to a recursive approach, to abstract versions of the counterterm character in general Rota--Baxter and dendriform algebras. Our approach allows us to understand the recursive nature of renormalization on the level of the pro-nilpotent Lie algebra of Feynman graphs. It turns out that the Baker--Campbell--Hausdorff recursion respectively the pre-Lie Magnus expansion provide the Lie algebraic analog of Bogoliubov's preparation map for Feynman graphs. We finish this survey by giving a short account of a natural matrix setting for perturbative renormalization which is well-suited for low-order explicit computations. Moreover, it allows for the transparent realization of the aforementioned abstract findings. 

\medskip

Let us briefly outline the organization of this survey. In the next section we review the essential notions from Hopf algebra theory. This section finishes with an abstract review on Connes--Kreimer's Birkhoff decomposition of Feynman rules in terms of Bogoliubov's preparation map. Subsection~\ref{subsect:BCHrecursion} contains an approach to Connes--Kreimer's Birkhoff decomposition based on a recursion defined in terms of the Baker--Campbell--Hausdorff formula. In section~\ref{sect:RB} we analyze Bogoliubov's recursion from the point of view of Rota--Baxter algebras. These algebras provide the natural tools to understand Connes--Kreimer's finding of a factorization solved by Bogoliubov's formula. As it turns out, Loday's dendriform algebras serve as an abstract algebraic frame for one of the main aspects, i.e. iteration of Rota--Baxter maps and a particular interplay between an associative and a pre-Lie product induced by the Rota--Baxter structure. This section ends with a continuation of the last subsection of section~\ref{sect:RB} dwelling on aspects related to renormalization theory. Finally, we briefly mention a non-Hopfian approach to Connes--Kreimer's finding in terms of triangular matrices providing a simple and straightforward setting for renormalization.

\section{A summary of Birkhoff--Connes--Kreimer factorization}
\label{sect:BirkhoffCKdecomposition}

We introduce the crucial property of connectedness for bialgebras. The main interest resides in the possibility to implement recursive procedures in connected bialgebras, the induction taking place with respect to a filtration (e.g. the coradical filtration) or a grading. An important example of these techniques is the recursive construction of the antipode, which then ``comes for free'', showing that any connected bialgebra is in fact a connected Hopf algebra. The recursive nature of Bogoliubov's formula in the BPHZ \cite{BP,H,Z}  approach to perturbative renormalization ultimately comes from the connectedness of the underlying Hopf algebra respectively the corresponding pro-nilpotency of the Lie algebra of infinitesimal characters.

For details on bialgebras and Hopf algebras we refer the reader to the standard references, e.g. \cite{Sw}. The use of bialgebras and Hopf algebras in combinatorics can at least be traced back to the seminal work of Joni and Rota~\cite{JR}.

\subsection{Connected graded bialgebras}
 
Let $k$ be a field with characteristic zero. A {\sl graded Hopf algebra\/} on $k$ is a graded $k$-vector space:
$$
	\mathcal H = \bigoplus_{n\ge 0}\mathcal H_n
$$
endowed with a product $m:\mathcal H \otimes \mathcal H\to\mathcal H$, a coproduct
$\Delta: \mathcal H \to \mathcal H \otimes \mathcal H$, a unit $u: k \to \mathcal H$, 
a co-unit $\varepsilon: \mathcal H \to k$ and an antipode $S: \mathcal H \to \mathcal H$ 
fulfilling the usual axioms of a Hopf algebra~\cite{Sw}, and such that:
 \allowdisplaybreaks{
\begin{align*}
	m(\mathcal H_p \otimes \mathcal H_q)	&\subset \mathcal H_{p+q},	\\
	\Delta(\mathcal H_n)	            &\subset \bigoplus_{p+q=n}\mathcal H_p\otimes\mathcal H_q,\\
	S(\mathcal H_n)	                  &\subset \mathcal H_n.
\end{align*}}
If we do not ask for the existence of an antipode $S$ on $\mathcal H$ we get the definition 
of a {\sl graded bialgebra\/}. In a graded bialgebra $\mathcal H$ we shall consider the 
increasing filtration:
$$
	\mathcal H^n = \bigoplus_{p=0}^n\mathcal H_p.
$$
Suppose moreover that $\mathcal H$ is {\sl connected\/}, i.e. $\mathcal H_{0}$ is one-dimensional. 
Then we have: 
$$
	\mop{Ker}\varepsilon=\bigoplus_{n\ge 1}\mathcal H_n.
$$

\begin{proposition}\label{rec1}
For any $x\in\mathcal H^n, n\ge 1$ we can write:
$$
	\Delta x = x\otimes\un + \un\otimes x + \widetilde\Delta x,
	\hskip 12mm 
	\widetilde\Delta x \in \bigoplus_{{p+q=n,}\atop {p\not= 0, q\not= 0}}\mathcal H_p\otimes \mathcal H_q.
$$
The map $\widetilde\Delta$ is coassociative on $\mop{Ker}\varepsilon$ and $\widetilde\Delta_k:=(I^{\otimes k-1} \otimes \widetilde \Delta)(I^{\otimes k-2} \otimes \widetilde\Delta) \cdots \widetilde\Delta$ sends $\mathcal H^n$ into $(\mathcal H^{n-k})^{\otimes k+1}$.
\end{proposition}

\begin{proof}
Thanks to connectedness we clearly can write:
$$
	\Delta x = a(x\otimes \un) + b(\un \otimes x) + \widetilde\Delta x
$$
with $a,b\in k$ and $\widetilde\Delta x\in\mop{Ker}\varepsilon\otimes\mop{Ker}\varepsilon$. 
The co-unit property then tells us that, with $k\otimes \mathcal H$ and $\mathcal H\otimes k$ 
canonically identified with $\mathcal H$:
\begin{equation*}
	x	=(\varepsilon\otimes I)(\Delta x)=bx,
	\hskip 15mm	
	x	=(I\otimes\varepsilon)(\Delta x)=ax,
\end{equation*}
hence $a=b=1$. We shall use the following two variants of Sweedler's notation:
 \allowdisplaybreaks{
\begin{align*}
	\Delta x	  = \sum_{(x)}x_1\otimes x_2,	\qquad\
	\widetilde\Delta x	= \sum_{(x)}x'\otimes x'',
\end{align*}}
the second being relevant only for $x\in\mop{Ker}\varepsilon$. If $x$ is homogeneous of degree $n$ we can suppose that the components $x_1,x_2,x'$, and $x''$ in the expressions above are homogeneous as well, and we have then $|x_1|+|x_2|=n$ and $|x'|+|x''|=n$. We easily compute:
 \allowdisplaybreaks{
\begin{align*}
	(\Delta\otimes I)\Delta(x)	&= x\otimes \un \otimes \un  + 
	                                  \un \otimes x\otimes \un  + 
	                                    \un \otimes \un \otimes x	\\
				                      &\ + \sum_{(x)} x'\otimes x''\otimes \un  
				                              + x'\otimes \un \otimes x'' 
				                               + \un \otimes x'\otimes x''\\
				                      &\ +(\widetilde\Delta\otimes I)\widetilde \Delta (x)
\end{align*}}
and
 \allowdisplaybreaks{
\begin{align*}
	(I\otimes\Delta)\Delta(x)	&=x\otimes \un \otimes \un  
	                              +\un \otimes x\otimes \un 
	                      					+\un \otimes 1\otimes x	\\
				                    &\ +\sum_{(x)} x'\otimes x''\otimes \un  
				                        + x'\otimes \un \otimes x'' 
				                         + \un \otimes x'\otimes x''\\
				                    &\ +(I\otimes\widetilde\Delta)\widetilde \Delta (x),
\end{align*}}
hence the co-associativity of $\widetilde\Delta$ comes from the one of $\Delta$. Finally it is easily seen by induction on $k$ that for any $x\in\mathcal H^n$ we can write:
\begin{equation*}
	\widetilde\Delta_k(x)=\sum_{x}x^{(1)}\otimes\cdots\otimes x^{(k+1)},
\end{equation*}
with $|x^{(j)}|\ge 1$. The grading imposes:
$$
	\sum_{j=1}^{k+1}|x^{(j)}|=n,
$$
so the maximum possible for any degree $|x^{(j)}|$ is $n-k$.
\end{proof}

\subsection{Connected filtered bialgebras}

A {\sl filtered Hopf algebra\/} on $k$ is a $k$-vector space together with an increasing 
$\mathbb{Z}_+$-indexed filtration: 
$$
	\mathcal H^0\subset\mathcal H^1\subset\cdots\subset \mathcal H^n\subset\cdots, \;\;\;\bigcup_n \mathcal H^n=\mathcal H
$$
endowed with a product $m:\mathcal H\otimes \mathcal H\to\mathcal H$, a coproduct $\Delta:\mathcal H\to\mathcal H\otimes \mathcal H$, a unit $u:k\to\mathcal H$, a co-unit $\varepsilon:\mathcal H\to k$ and an antipode $S:\mathcal H\to\mathcal H$
fulfilling the usual axioms of a Hopf algebra, and such that:
 \allowdisplaybreaks{
\begin{align*}
	m(\mathcal H^p \otimes \mathcal H^q)	\subset \mathcal H^{p+q},	
	\qquad
	\Delta(\mathcal H^n)	            \subset \sum_{p+q=n}\mathcal H^p\otimes\mathcal H^q, 
	\;\;\;\;\; {\rm{ and }}\; 
	S(\mathcal H^n)	                  \subset \mathcal H^n.
\end{align*}}
If we do not ask for the existence of an antipode $S$ on $\mathcal H$ we get the definition of a {\sl filtered bialgebra\/}. For any $x\in\mathcal H$ we set:
\begin{equation*}
	|x|:=\mop{min}\{n\in\mathbb{N},\ x\in\mathcal H^n\}.
\end{equation*}
Any graded bialgebra or Hopf algebra is obviously filtered by the canonical filtration associated to the grading:
\begin{equation*}
	\mathcal H^n:=\bigoplus_{i=0}^n \mathcal H_i,
\end{equation*}
and in that case, if $x$ is an homogeneous element, $x$ is of degree $n$ if and only if $|x|=n$. We say that the filtered bialgebra $\mathcal H$ is connected if $\mathcal H^0$ is one-dimensional. There is an analogue of Proposition \ref{rec1} in the connected filtered case, the proof of which is very similar:

\begin{proposition}
For any $x\in\mathcal H^n, n\ge 1$ we can write:
\begin{equation*}
	\Delta x=x\otimes\un+\un\otimes x+\widetilde\Delta x,
	\hskip 15mm 
	\widetilde\Delta x\in\sum_{{p+q=n,} \atop {p\not= 0, q\not= 0}}\mathcal H^p\otimes\mathcal H^q.
\end{equation*}
The map $\widetilde\Delta$ is coassociative on $\mop{Ker}\varepsilon$ and $\widetilde\Delta_k=(I^{\otimes k-1}\otimes\widetilde\Delta)(I^{\otimes k-2}\otimes\widetilde\Delta)\cdots \widetilde\Delta$ sends $\mathcal H^n$ into $(\mathcal H^{n-k})^{\otimes k+1}$.
\end{proposition}

\noindent
The coradical filtration endows any pointed Hopf algebra $\mathcal H$ with a structure of filtered Hopf algebra (S. Montgomery, \cite{Mo} Lemma 1.1). If $\mathcal H$ is moreover irreducible (i.e. if the image of $k$ under the unit map $u$ is the unique one-dimensional simple subcoalgebra of $\mathcal H$) this filtered Hopf algebra is moreover connected.

\subsection{The convolution product}
\label{convolution}

An important result is that any connected filtered bialgebra is indeed a filtered Hopf algebra, in the sense that the antipode comes for free. We give a proof of this fact as well as a recursive formula for the antipode with the help of the {\sl convolution product\/}: let $\mathcal H$ be a (connected filtered)  bialgebra, and let $\mathcal A$ be any $k$-algebra (which will be called the 
{\sl target algebra\/}). The convolution product on the space $\mathcal L(\mathcal H,\mathcal A)$ of linear maps from $\mathcal{H}$ to $\mathcal{A}$ is 
given by:
 \allowdisplaybreaks{
\begin{align*}
	\varphi*\psi (x)	&=m_{\mathcal A}(\varphi\otimes\psi)\Delta(x)\\
			 	&=\sum_{(x)}\varphi(x_1)\psi(x_2).
\end{align*}}

\begin{proposition}\label{prop:groupeG}
The map $e=u_{\mathcal A}\circ\varepsilon$, given by $e(\un)=\un_{\mathcal A}$ and $e(x)=0$ for any $x \in \mop{Ker} \varepsilon$, is a unit for the convolution product. Moreover the set $G(\mathcal A):=\{\varphi \in \mathcal L(\mathcal H,\mathcal A),\ \varphi(\un)=\un_{\mathcal A}\}$ endowed with the convolution product is a group.
\end{proposition}

\begin{proof}
The first statement is straightforward. To prove the second let us consider the formal series:
 \allowdisplaybreaks{
\begin{align*}
	\varphi^{*-1}(x)	&=\bigl(e-(e-\varphi)\bigr)^{*-1}(x)	\\
				&=\sum_{m\ge 0}(e-\varphi)^{*m}(x).
\end{align*}}
Using $(e-\varphi)(\un)=0$ we have immediately $(e-\varphi)^{*m}(\un)=0$, and
for any $x\in\mop{Ker}\varepsilon$:
\begin{equation*}
	(e-\varphi)^{*n}(x)=m_{\mathcal A,n-1}
	(\underbrace{\varphi\otimes\cdots\otimes\varphi}_{n\hbox{ \sevenrm times }})\widetilde\Delta_{n-1}(x).
\end{equation*}
When $x\in\mathcal H^p$ this expression vanishes then for $n\ge p+1$. The formal series ends up then with a finite number of terms for any $x$, which proves the result.
\end{proof}

\begin{corollary}\label{antipode}
Any connected filtered bialgebra $\mathcal H$ is a filtered Hopf algebra. The antipode is defined by:
\begin{equation}
\label{eq:antipode}
	S(x)=\sum_{m\ge 0}(u\circ\varepsilon-I)^{*m}(x).
\end{equation}
It is given by $S(\un)=\un$ and recursively by any of the two formulas for $x \in \mop{Ker}\varepsilon$:
 \allowdisplaybreaks{
\begin{align*}
			S(x)	=-x-\sum_{(x)}S(x')x''			
			\;\;\; \ {\rm{ and }}\;\;\; \
			S(x)	=-x-\sum_{(x)}x'S(x'').	  
\end{align*}}
\end{corollary}

\begin{proof}
The antipode, when it exists, is the inverse of the identity for the convolution product on 
$\mathcal L(\mathcal H,\mathcal H)$. One just needs then to apply Proposition \ref{prop:groupeG} with 
$\mathcal A=\mathcal H$. The two recursive formulas follow directly from the two equalities:
$$
	m(S\otimes I)\Delta (x)=\ 0 \ =m(I\otimes S)\Delta (x)
$$
fulfilled by any $x\in\mop{Ker}\varepsilon$.
\end{proof}

Let $\mathfrak g(\mathcal A)$ be the subspace of $\mathcal L(\mathcal H,\mathcal A)$ formed by the elements 
$\alpha$ such that $\alpha(\un)=0$. It is clearly a subalgebra of $\mathcal L(\mathcal H,\mathcal A)$ 
for the convolution product. We have:
\begin{equation}
\label{groupG}
	G(\mathcal A) = e + \mathfrak g(\mathcal A).
\end{equation}
From now on we shall suppose that the ground field $k$ is of characteristic zero. For any $x\in\mathcal H^n$ the exponential:
\begin{equation*}
	\exp^*(\alpha)(x)=\sum_{k\ge 0}{\alpha^{*k}(x)\over k!}
\end{equation*}
is a finite sum (ending up at $k=n$). It is a bijection from $\mathfrak g(\mathcal A)$
onto $G(\mathcal A)$. Its inverse is given by:
\begin{equation*}
	\log^*(e+\alpha)(x)=\sum_{k\ge 1}{(-1)^{k-1}\over k}\alpha^{*k}(x).
\end{equation*}
This sum again ends up at $k=n$ for any $x\in\mathcal H^n$. Let us introduce a decreasing filtration on $\mathcal L=\mathcal L(\mathcal H,\mathcal A)$:
\begin{equation*}
	\mathcal L^n:=\{\alpha\in\mathcal L, \ \alpha\restr{\mathcal H^{n-1}}=0\}.
\end{equation*}
Clearly $\mathcal L^0=\mathcal L$ and $\mathcal L^1=\mathfrak g(\mathcal A)$. We define the valuation $\mop{val}\varphi$ of an element $\varphi$ of $\mathcal L$ as the biggest integer $k$ such that $\varphi$ is in $\mathcal L^k$. We shall consider in the sequel the ultrametric distance on $\mathcal L$ induced by the filtration:
\begin{equation}
\label{distance}
	d(\varphi,\psi)=2^{-\smop{val}(\varphi-\psi)}.
\end{equation}
For any $\alpha,\beta\in\mathfrak g(\mathcal A)$ let $[\alpha,\beta]=\alpha*\beta-\beta*\alpha$.

\begin{proposition}
We have the inclusion:
\begin{equation*}
	\mathcal L^p*\mathcal L^q\subset\mathcal L^{p+q},
\end{equation*}
and moreover the metric space $\mathcal L$ endowed with the distance defined by (\ref{distance}) is complete.
\end{proposition}

\begin{proof}
Take any $x\in\mathcal H^{p+q-1}$, and any $\alpha\in\mathcal L^p$ and $\beta\in\mathcal L^q$. We have
$$
	(\alpha*\beta)(x)=\sum_{(x)}\alpha(x_1)\beta(x_2).
$$
Recall that we denote by $|x|$ the minimal $n$ such that $x\in\mathcal H^n$. Since $|x_1|+|x_2|=|x|\le p+q-1$, either $|x_1|\le p-1$ or $|x_2|\le q-1$, so the expression vanishes. Now if $(\psi_n)$ is a Cauchy sequence in $\mathcal L$ it is immediate to see that this sequence is {\sl locally stationary\/}, i.e. for any $x\in\mathcal H$ there exists $N(x)\in\mathbb{N}$ such that $\psi_n(x)=\psi_{N(x)}(x)$ for any $n\ge N(x)$. Then the limit of $(\psi_n)$ exists and is clearly defined by:
$$
	\psi(x)=\psi_{N(x)}(x).
$$  
\end{proof}

As a corollary the Lie algebra $\mathcal L^1=\mathfrak g(\mathcal A)$ is {\sl pro-nilpotent}, in a sense that it is the projective limit of the Lie algebras $\mathfrak g(\mathcal A)/\mathcal L^n$, which are nilpotent.

\subsection{Characters and infinitesimal characters}
\label{sect:characters}

Let $\mathcal H$ be a connected filtered Hopf algebra over $k$, and let $\mathcal A$ be a {\sl commutative\/} $k$-algebra. 
We shall consider unital algebra morphisms from $\mathcal H$ to the target algebra $\mathcal A$, which we shall call slightly abusively {\sl characters\/}. We recover of course the usual notion of character when the algebra $\mathcal A$ is the ground field $k$. The notion of character involves only the algebra structure of $\mathcal H$. On the other hand the convolution product on $\mathcal L(\mathcal H,\mathcal A)$ involves only the {\sl coalgebra\/} structure on $\mathcal H$. Let us consider now the full Hopf algebra structure on $\mathcal H$ and see what happens to characters with the convolution product:

\begin{proposition}\label{prop:convolution2}
Let $\mathcal H$ be a connected filtered Hopf algebra over $k$, and let $\mathcal A$ be a commutative $k$-algebra. Then the characters from $\mathcal H$ to $\mathcal A$ form a group $G_1(\mathcal A)$ under the convolution product, and for any $\varphi\in G_1(\mathcal A)$ the inverse is given by:
\begin{equation*}
	\varphi^{*-1}=\varphi\circ S.
\end{equation*}
\end{proposition}
\noindent

We call {\sl  infinitesimal characters with values in the algebra $\mathcal A$\/} those elements $\alpha$ of $\mathcal L(\mathcal H,\mathcal A)$ such that:
$$
	\alpha(xy)=e(x)\alpha(y)+\alpha(x)e(y).
$$

\begin{proposition}\label{prop:exp}
Let $G_1(\mathcal A)$ (resp. $\mathfrak g_1(\mathcal A)$) be the set of characters of $\mathcal H$ with values in
$\mathcal A$ (resp. the set of infinitesimal characters of $\mathcal H$ with values in $\mathcal A$). Then 
$G_1(\mathcal A)$ is a subgroup of $G$, the exponential restricts to a bijection from $\mathfrak g_1(\mathcal A)$ 
onto $G_1(\mathcal A)$, and $\mathfrak g_1(\mathcal A)$ is a Lie subalgebra of $\mathfrak g(\mathcal A)$.
\end{proposition}

\begin{proof}
Part of these results are a reformulation of Proposition \ref{prop:convolution2} and some
points are straightforward. The only non-trivial point concerns $\mathfrak g_1(\mathcal A)$ and
$G_1(\mathcal A)$. Take two infinitesimal characters $\alpha$ and $\beta$ with values in
$\mathcal A$ and compute:
 \allowdisplaybreaks{
\begin{align*}
	(\alpha*\beta)(xy)
	&=\sum_{(x)(y)}\alpha(x_1y_1)\beta(x_2y_2)	\\
	&=\sum_{(x)(y)}\bigl(\alpha(x_1)e(y_1)+e(x_1)\alpha(y_1)\bigr).
									\bigl(\beta(x_2)e(y_2)+e(x_2)\alpha(y_2)\bigr)		\\
	&=(\alpha*\beta)(x)e(y)+\alpha(x)\beta(y)+\beta(x)\alpha(y) + e(x)(\alpha*\beta)(y).
\end{align*}}
Using the commutativity of $\mathcal A$ we immediately get:
$$
	[\alpha,\beta](xy)=[\alpha,\beta](x)e(y)+e(x)[\alpha,\beta](y),
$$
which shows that $\mathfrak g_1(\mathcal A)$ is a Lie algebra. Now for $\alpha\in\mathfrak g_1(\mathcal A)$ we have:
$$
	\alpha^{*n}(xy)=\sum_{k=0}^n{n\choose k}\alpha^{*k}(x)\alpha^{*(n-k)}(y),
$$
as easily seen by induction on $n$. A straightforward computation then yields:
$$
	\exp^*(\alpha)(xy)=\exp^*(\alpha)(x)\exp^*(\alpha)(y).
$$
\end{proof}

\subsection{Renormalisation in connected filtered Hopf algebras}
\label{sect:ren}

We describe in this section the renormalisation \`a la Connes--Kreimer (\cite{kreimer1998}, \cite{CK}, \cite{CK1}) in the abstract context of connected filtered Hopf algebras: the objects to be renormalised are characters 
with values in a commutative unital target algebra $\mathcal A$ endowed with a {\sl renormalisation scheme\/}, i.e. a splitting $\mathcal A=\mathcal A_-\oplus\mathcal A_+$ into two subalgebras with $\un\in\mathcal A_+$. An important example is given by the {\sl minimal subtraction\/} (MS) {\sl scheme\/} on the algebra $\mathcal A$ of meromorphic functions of one variable $z$, where $\mathcal A_+$ is the algebra of meromorphic functions which are holomorphic at $z=0$, and where $\mathcal A_-=z^{-1}\mathbb{C}[z^{-1}]$ stands for the ``polar parts''. Any $\mathcal A$-valued character $\varphi$ admits a unique {\sl Birkhoff decomposition\/}
$$
	\varphi=\varphi_-^{*-1}*\varphi_+,
$$
where $\varphi_+$ is an $\mathcal A_+$-valued character, and where $\varphi_-(\mop{Ker}\varepsilon)\subset \mathcal A_-$. In the MS scheme case described just above, the renormalised character is the scalar-valued character given by the evaluation of $\varphi_+$ at $z=0$ (whereas the evaluation of $\varphi$ at $z=0$ does not necessarily make sense).

\begin{theorem} \label{th:main} Factorization of the group $G(\mathcal A)$ 
\begin{enumerate}
	\item Let $\mathcal H$ be a connected filtered Hopf algebra. Let $\mathcal A$ be a commutative unital algebra with a renormalisation scheme, and let $\pi:\mathcal A \to \mathcal A$ be the projection onto $\mathcal A_-$ 				parallel to $\mathcal A_+$. Let $G(\mathcal A)$ be the group of those $\varphi\in\mathcal L(\mathcal H,\mathcal A)$ such 		that $\varphi(\un)=\un_{\mathcal A}$ endowed with the convolution product. Any $\varphi\in G(\mathcal A)$ admits a unique Birkhoff decomposition
\begin{equation}
\label{Birkhoff}
	\varphi=\varphi_-^{*-1} * \varphi_+,
\end{equation}
where $\varphi_-$ sends $\un$ to $\un_{\mathcal A}$ and $\mop{Ker}\varepsilon$ into $\mathcal A_-$, and where $\varphi_+$ sends $\mathcal H$ into $\mathcal A_+$. The maps $\varphi_-$ and $\varphi_+$ are given on $\mop{Ker}\varepsilon$ by the following recursive formulas
 \allowdisplaybreaks{
\begin{align*}
	\varphi_-(x)	&=-\pi\Bigl( \varphi(x)+\sum_{(x)}\varphi_-(x')\varphi(x'')\Bigr)	\\
	\varphi_+(x)	&=(I-\pi)\Bigl(\varphi(x)+\sum_{(x)}\varphi_-(x')\varphi(x'')\Bigr).
\end{align*}}
where $I$ is the identity map.
	
	\item If $\varphi$ is a character, the components $\varphi_-$ and $\varphi_+$ occurring in the Birkhoff decomposition of $\varphi$ are characters as well. 
\end{enumerate}
\end{theorem}

\begin{proof}
The proof goes along the same lines as the proof of Theorem 4 of \cite{CK1}: for the first assertion it is immediate from the definition of $\pi$ that $\varphi_-$ sends $\mop{Ker}\varepsilon$ into $\mathcal A_-$, and that $\varphi_+$ sends $\mop{Ker}\varepsilon$ into $\mathcal A_+$. It only remains to check equality $\varphi_+=\varphi_-*\varphi$, which is an easy computation
 \allowdisplaybreaks{
\begin{align*}
\varphi_+(x)	&=(I-\pi)\Bigl( \varphi(x)+\sum_{(x)}\varphi_-(x')\varphi(x'')\Bigr).\\
			&=\varphi(x)+\varphi_-(x)+ \sum_{(x)}\varphi_-(x')\varphi(x'')	\\
			&=(\varphi_-*\varphi)(x).
\end{align*}}
The proof of assertion 2) can be carried out exactly as in \cite{CK1} and relies on the
following {\sl Rota--Baxter relation\/} in $\mathcal A$:
\begin{equation}
\label{RBrelation1}
	\pi(a)\pi(b)=\pi\bigl(\pi(a)b+a\pi(b)\bigr)-\pi(ab),
\end{equation}
which is easily verified by decomposing $a$ and $b$ into their $\mathcal A_\pm$-parts. We will derive below a more conceptual proof.
\end{proof}

\begin{remark}
Define the {\rm Bogoliubov preparation map\/} as the map $B: G(\mathcal A)\to \mathcal L(\mathcal H,\mathcal A)$ given by:
\begin{equation}
\label{bogo3}
	B(\varphi) = \varphi_- * (\varphi - e),
\end{equation}
such that for any $x \in \mop{Ker}\varepsilon$ we have:
\begin{equation*}
	B(\varphi)(x)=\varphi(x) + \sum_{(x)}\varphi_-(x')\varphi(x'').
\end{equation*}
The components of $\varphi$ in the Birkhoff decomposition read:
\begin{equation}
\label{bogo2}
	\varphi_- = e - \pi\circ B(\varphi),
	\hskip 12mm 
	\varphi_+ = e + (I-\pi)\circ B(\varphi).
\end{equation}
On $\mop{Ker}\varepsilon$ they reduce to $-\pi\circ B(\varphi)$, $(I-\pi)\circ B(\varphi)$, respectively. Plugging equation (\ref{bogo3}) inside (\ref{bogo2}) and setting $\alpha:=e-\varphi$ we get the following expression for $\varphi_-$:
 \allowdisplaybreaks{
\begin{align}
	\varphi_-	&=	e+P(\varphi_-*\alpha) 							\label{pre-spitzer-} \\
			&=	e+P(\alpha)+P\bigl(P(\alpha) * \alpha\bigr)+ \cdots +
					\underbrace{P\Bigl(P\bigl(\ldots P(}_{\hbox{\sevenrm $\scriptstyle n$ times}}
					\alpha)*\alpha\bigr)\cdots *\alpha\Bigr)+ \cdots     \nonumber   
\end{align}}
and for $\varphi_+$ we find:
 \allowdisplaybreaks{
\begin{align}
	\varphi_+	&=	e-\tilde{P}(\varphi_-*\alpha) 						\label{pre-spitzer+} \\
				&= e-\tilde P(\varphi_+*\beta)\\
			&=	e-\tilde{P}(\beta)+\tilde{P}\bigl(\tilde{P}(\beta) * \beta\bigr) - \cdots +
				(-1)^n \underbrace{\tilde{P}\Bigl(\tilde{P}\bigl(\ldots \tilde{P}(}_{\hbox{\sevenrm $\scriptstyle n$ times}} 	  
								\beta)*\beta\bigr)\cdots *\beta\Bigr)+ \cdots     \nonumber   
\end{align}}
with $\beta:=\varphi^{-1}*\alpha=\varphi^{-1}-e$, and where $\tilde{P}$ and $P$ are projections on $\mathcal L(\mathcal H,\mathcal A)$ defined by $\tilde{P}(\alpha)=(I-\pi) \circ \alpha$ and $P(\alpha)=\pi \circ \alpha$, 
respectively.
\end{remark}

\subsection{The Baker--Campbell--Hausdorff recursion}
\label{subsect:BCHrecursion}

Let $\mathcal L$ be any complete filtered Lie algebra. Thus $\mathcal L$ has a
decreasing filtration $(\mathcal L_n)$ of Lie subalgebras such that $[\mathcal L_m,\,\mathcal L_n]
\subseteq \mathcal L_{m+n}$ and $\mathcal L \cong \mopl{lim}_{\leftarrow} \mathcal L/\mathcal L_n$ 
(i.e., $\mathcal L$ is complete with respect to the topology induced by the filtration). 
Let $A$ be the completion of the enveloping algebra $\mathcal U(\mathcal L)$ for the decreasing
filtration naturally coming from that of $\mathcal L$. The functions:
 \allowdisplaybreaks{
\begin{align*}
    \exp: A_1 \to 1+A_1&, \qquad \exp(a)=\sum_{n=0}^\infty \frac{a^n}{n!},\\
    \log: 1+A_1 \to A_1&, \qquad \log (1+a)=-\sum_{n=1}^\infty \frac{(-a)^n}{n}
\end{align*}
are well-defined and are the inverse of each other. The
Baker--Campbell--Hausdorff ($BC\!H$) formula writes for any $x,y\in\mathcal L_1$~\cite{R,V}:
\begin{equation*}
    \exp(x)\exp(y) = \exp\bigl(C(x,y)\bigr) = \exp\bigl(x+y+\mop{BCH}(x,y)\bigr),
\end{equation*}
where $\mop{BCH}(x,y)$ is an element of $\mathcal L_2$ given by a Lie series the first few terms of 
which are:
\begin{equation*}
    \mop{BCH}(x,y)=\frac{1}{2}[x,y]+\frac{1}{12}[x,[x,y]]+\frac{1}{12}[y,[y,x]]-
	                \frac{1}{24}[x,[y,[x,y]]] +\cdots
\end{equation*}
Now let $P: \mathcal L \to \mathcal L$ be any linear map preserving the filtration
of $\mathcal L$. We define $\tilde{P}$ to be $\mop{Id}_{\mathcal L}-P$. For $a \in \mathcal L_1$,
define $\chi(a) = \lim_{n \to \infty} \chi_{(n)}(a)$ where $\chi_{(n)}(a)$ is given by the $BC\!H$-recursion:
 \allowdisplaybreaks{
\begin{align}
    \chi_{(0)}(a) 	  &:= a, \nonumber\\
    \chi_{(n+1)}(a) &= a - 
    \mop{BCH}\bigl( P(\chi_{(n)}(a)),\,(\mop{Id}_{\mathcal L}-P)(\chi_{(n)}(a)) \bigr),
    \label{eq:chik}
\end{align}}
and where the limit is taken with respect to the topology given by
the filtration. Then the map $\chi: \mathcal L_1 \to \mathcal L_1$ satisfies:
\begin{equation}
    \label{BCHrecursion1}
    \chi(a) = a - \mop{BCH}\bigl(P(\chi(a)),\,\tilde{P}(\chi(a))\bigr).
\end{equation}
This map appeared in \cite{EGK2}, \cite{EGK3}, where more details can be found, see also \cite{M2,M3}. The following proposition (\cite{EGM}, \cite{M2}) gives further properties of the map $\chi$.

\begin{proposition} \label{prop:conv}
For any linear map $P: \mathcal L \to \mathcal L$ preserving the filtration of $\mathcal L$
there exists a (usually non-linear) unique map $\chi: \mathcal L_1 \to \mathcal L_1$
such that $(\chi - \mop{Id}_{\mathcal L})(\mathcal L_i) \subset \mathcal L_{2i}$ for any $i \ge 1$,
and such that, with $\tilde P:= \mop{Id}_{\mathcal L} - P$ we have:
\begin{equation}
    \label{truc}
    \forall a \in \mathcal L_1,\ \ a = C \Bigl(P\bigl(\chi(a)\bigr),\,\tilde P\bigl(\chi(a)\bigr)\Bigr).
\end{equation}
This map is bijective, and its inverse is given by:
\begin{equation}
    \label{chi-inverse}
    \chi^{-1}(a) = C\bigl(P(a),\,\tilde{P}(a)\bigr)
                 = a+\mop{BCH}\bigl(P(a),\,\tilde P(a)\bigr).
\end{equation}
\end{proposition}

\begin{proof}
Equation (\ref{truc}) can be rewritten as:
$$
    \chi(a)=F_a \bigl(\chi(a)\bigr),
$$
with $F_a: \mathcal L_1 \to \mathcal L_1$ defined by:
$$
    F_a(b)=a-\mop{BCH}\bigl(P(b),\tilde{P}(b)\bigr).
$$
This map $F_a$ is a contraction with respect to the metric
associated with the filtration: indeed if $ b, \varepsilon \in \mathcal L_1$
with $\varepsilon \in \mathcal L_n$, we have:
$$
    F_a(b + \varepsilon) - F_a(b) = \mop{BCH}\bigl(P(b),\,\tilde{P}(b)\bigr)                    
    -\mop{BCH}\bigl(P(b+\varepsilon),\,\tilde{P}(b+\varepsilon)\bigr).
$$
The right-hand side is a sum of iterated commutators in each of which $\varepsilon$ does appear at least once. So it belongs to
$\mathcal L_{n+1}$. So the sequence $F_a^n(b)$ converges in $\mathcal L_1$ to a unique fixed point $\chi(a)$ for $F_a$.

\medskip

Let us remark that for any $a \in \mathcal L_i$, then, by a straightforward induction argument, $\chi_{(n)}(a)\in \mathcal L_i$ for any $n$, so $\chi(a)\in \mathcal L_i$ by taking the limit. Then the difference $\chi(a)-a=\mop{BCH}\bigl(P\bigl(\chi(a)\bigr),\, \tilde P\bigl(\chi(a)\bigr)\bigr)$ clearly belongs to $\mathcal L_{2i}$. Now consider the map $\psi:\mathcal L_1\to \mathcal L_1$ defined by $\psi(a)=C\bigl(P(a),\,\tilde P(a)\bigr)$. It is clear from the definition of $\chi$ that $\psi\circ\chi=\mop{Id}_{\mathcal L_1}$. Then $\chi$ is injective and $\psi$ is surjective. The injectivity of $\psi$ will be an immediate consequence of the following lemma.

\begin{lemma}\label{lem:dilation}
	The map $\psi$ increases the ultrametric distance given by the filtration.
\end{lemma}

\begin{proof}
For any $x,y \in \mathcal L_1$ the distance $d(x,y)$ is given by $2^{-n}$ where $n=\hbox{sup}\{k\in\mathbb{N},\,x-y\in \mathcal L_k\}$. We have then to prove that $\psi(x)-\psi(y)\notin \mathcal L_{n+1}$. But:
 \allowdisplaybreaks{
\begin{align*}
  \lefteqn{\psi(x)-\psi(y) = x-y+\mop{BCH}\bigl(P(x),\,\tilde P(x)\bigr)-\mop{BCH}\bigl(P(y),\,\tilde{P}(y)\bigr)}\\
                          &= x-y+\Bigl(\mop{BCH}\bigl(P(x),\,\tilde{P}(x)\bigr)
                        - \mop{BCH}\bigl(P(x)-P(x-y),\,\tilde{P}(x) - \tilde{P}(x-y)\bigr)\Bigr).
\end{align*}}
The rightmost term inside the large brackets clearly belongs to $\mathcal L_{n+1}$. As $x-y\notin \mathcal L_{n+1}$ by hypothesis, this proves the claim.
\end{proof}

\noindent
The map $\psi$ is then a bijection, so $\chi$ is also bijective, which proves Proposition \ref{prop:conv}.
\end{proof}

\begin{corollary} \label{cor:birkhoffBCH}
For any $a\in\mathcal L_1$ we have the following equality taking place in $1+A_1\subset A$:
\begin{equation}
\label{birkhoff2}
	\exp(a)=\exp\bigl(P(\chi(a))\bigr)\exp\bigl(\tilde{P}(\chi(a))\bigr).
\end{equation}
\end{corollary}
Putting (\ref{pre-spitzer-}) and (\ref{birkhoff2}) together we get for any $\alpha\in\mathcal L_1$ the following {\sl non-commutative Spitzer identity\/}:
\begin{equation}
\label{ncspitzer}
	e+P(\alpha)+
%	P\bigl(P(\alpha)*\alpha\bigr)+
                            	\cdots+
		\underbrace{P\Bigl(P\bigl(\ldots P(}_{\hbox{\sevenrm $\scriptstyle n$ times}}\alpha)
		*\alpha\bigr)\cdots *\alpha\Bigr)+\cdots=
		\exp\Big[-P\Bigl(\chi\bigl(\log(e-\alpha)\bigr)\Bigr)\Big].
\end{equation}
This identity is valid for any filtration-preserving Rota--Baxter operator $P$ in a complete filtered Lie algebra (see section \ref{sect:RB}). For a detailed treatment of these aspects, see \cite{EGK2}, \cite{EGK3}, \cite{EGM}, \cite{EMP}.

\begin{remark}
Using~(\ref{birkhoff2}) the reader should have no problem in verifying that the Baker--Campbell--Hausdorff recursion (\ref{BCHrecursion1}) can also be written more compactly: 
\begin{equation}
\label{BCHrecursion2}
    \chi(a) = a + \mop{BCH}\bigl(-P(\chi(a)),\ a\bigr).
\end{equation}
\end{remark}

\subsection{Application to perturbative renormalisation I}
\label{subsect: appl I}

Suppose now that $\mathcal L=\mathcal L(\mathcal H,\mathcal A)$ (with the setup and notations of paragraph \ref{sect:ren}), and that the operator $P$ is now the projection defined by $P(a)=\pi \circ a$. It is clear that Corollary~\ref{cor:birkhoffBCH} applies in this setting and that the first factor on the right-hand side of (\ref{birkhoff2}) is an element of $G_1(\mathcal A)$, the group of $\mathcal A$-valued characters of $\mathcal H$, which sends $\mop{Ker}\varepsilon$ into $\mathcal A_-$, and that the second factor is an element of $G_1$ which sends $\mathcal H$ into $\mathcal A_+$. Going back to Theorem \ref{th:main} and using uniqueness of the decomposition (\ref{Birkhoff}) we see then that (\ref{birkhoff2}) in fact is the Birkhoff--Connes--Kreimer decomposition of the element $\exp^*(a)$ in $G_1$. Indeed, starting with the infinitesimal character $a$ in the Lie algebra $\mathfrak g_1(\mathcal A)$ equation (\ref{birkhoff2}) gives the Birkhoff--Connes--Kreimer decomposition of $\varphi = \exp^*(a)$ in the group $G_1(\mathcal A)$ of $\mathcal A$-valued characters of $\mathcal H$, i.e.: 
$$
	\varphi_- = \exp^*\bigl(-P(\chi(a))\bigr)
	 \; \ {\rm{and}}\ \;
	\varphi_+ = \exp^*\bigl(\tilde{P}(\chi(a))\bigr)
	\ \; {\rm{such \ that}}\ \;
	\varphi = \varphi_-^{-1}* \varphi_+, 
$$ 
thus proving the second assertion in Theorem \ref{th:main}. 

Therefore, we may say that the Baker--Campbell--Hausdorff recursion (\ref{BCHrecursion2}) encodes the process of renormalization on the level of the Lie algebra $\mathfrak g_1(\mathcal A)$ of infinitesimal characters. Indeed, for $a \in \mathfrak g_1(\mathcal A)$, determined by the Feynman rules character $\varphi=\exp^*(a)$, we calculate the element $b(a):=\chi(a)$ in $\mathfrak g_1(\mathcal A)$, i.e. the Lie algebra analog of Bogoliubov's preparation map~(\ref{bogo3}), such that:  
\begin{equation}
\label{LieAlgRenrom}
	a = C\bigl(P(b(a)),\tilde{P}(b(a))\bigr) = C\bigl(P(\chi(a)),\tilde{P}(\chi(a)\bigr)   
\end{equation} 
gives rise to (\ref{birkhoff2}). By construction, $P(b(a))$ and $\tilde{P}(b(a))$ take values in 
$\mathcal{A}_{\mp}$, respectively. Hence, the decomposition (\ref{LieAlgRenrom}) of $a \in \mathfrak g_1(\mathcal A)$ is the Lie algebra analog of the Birkhoff--Connes--Kreimer decomposition of $\varphi = \exp^*(a)$.

Comparing Corollary~\ref{cor:birkhoffBCH} and Theorem \ref{th:main} the reader may wonder upon the role played by the Rota--Baxter relation~(\ref{RBrelation1}) for the projector $P$. In the following section we will show that it is this identity that allows to write the exponential $\varphi_- = \exp^*\bigl(-P(\chi(a))\bigr)$ as a recursion, that is, $\varphi_- = e - P(B(\varphi))$, where $B(\varphi)=\varphi_- * (\varphi-e)$. Equivalently, this amounts to the fact that the group $G_1(\mathcal A)$ factorizes into two subgroups $G_1^-(\mathcal A)$ and $G_1^+(\mathcal A)$, such that $\varphi_\pm \in G_1^\pm(\mathcal A)$.

\section{Rota--Baxter and dendriform algebras}
\label{sect:RB}

We are interested in abstract versions of identities (\ref{pre-spitzer-}) and (\ref{ncspitzer}) fulfilled by the counterterm character $\varphi_-$. The general algebraic context is given by Rota--Baxter (associative) algebras of weight $\theta$, which are themselves dendriform algebras. We first briefly recall the definition of Rota--Baxter (RB) algebra and its most important properties. For more details we refer the reader to the classical papers~\cite{A, B, Ca, Ro, RS}, as well as for instance to the references~\cite{EK, EGM}.

Let $A$ be an associative not necessarily unital nor commutative algebra with $R \in \mop{End}(A)$. The product of $a$ and $b$ in $A$ is written $a \cdot b$ or simply $ab$ when no confusion can arise. We call a tuple $(A,R)$ a {\textit{Rota--Baxter algebra}} of weight $\theta \in k$ if $R$ satisfies the {\textit{Rota--Baxter relation}}
\begin{equation}
    R(x)R(y) = R\bigl( R(x) y + x R(y) + \theta xy \bigr).
\label{def:RBR}
\end{equation}
Note that the operator $P$ of paragraph \ref{sect:ren} is an idempotent Rota--Baxter operator. Its weight is thus $\theta=-1$. Changing $R$ to $R' := \mu R$, $\mu \in k$, gives rise to a RB algebra of weight $\theta':=\mu \theta$, so that a change in the $\theta$ parameter can always be achieved, at least as long as weight non-zero RB algebras are considered. The definition generalizes to other types of algebras than associative algebras: for example one may want to consider RB Lie or pre-Lie algebra structures. Further below we will encounter examples of such structures. 

\medskip

Let us recall some classical examples of RB algebras. First, consider the integration by parts rule for the Riemann integral map. Let $A := C(\mathbb{R})$ be the ring of real continuous functions with pointwise product. The indefinite Riemann integral can be seen as a linear map on~$A$:
\begin{equation}
    I: A \to A, \qquad\  I(f)(x) := \int_0^x f(t)\,dt.
\label{eq:Riemann}
\end{equation}
Then, integration by parts for the Riemann integral can be written compactly as:
\begin{equation}
    I(f)(x)I(g)(x) = I\bigl( I(f) g \bigr)(x) + I\bigl( fI(g)\bigr)(x),
\label{eq:integ-by-parts}
\end{equation}
dually to the classical Leibniz rule for derivations. Hence, we found our first example of a weight zero Rota--Baxter map. Correspondingly, on a suitable class of functions, we define the following Riemann summation operators:
\begin{align}
    R_\theta(f)(x) := \sum_{n = 1}^{[x/\theta]} \theta f(n\theta)
    \qquad\ {\rm{and}} \qquad\
    R'_\theta(f)(x) := \sum_{n = 1}^{[x/\theta]+1} \theta f(n\theta).
\label{eq:le-clou}
\end{align}
We observe readily that:
 \allowdisplaybreaks{
\begin{align}
    &\biggl( \sum_{n = 1}^{[x/\theta]} \theta f(n\theta)\biggr)
     \biggl( \sum_{m = 1}^{[x/\theta]} \theta g(m\theta)\biggr)
   = \biggl( \sum_{n > m = 1}^{[x/\theta]}
        + \sum_{m > n = 1}^{[x/\theta]}
        + \sum_{m = n = 1}^{[x/\theta]}  \biggr)\theta^2 f(n\theta) g(m\theta) \nonumber \\
    &= \sum_{m = 1}^{[x/\theta]} \theta^2 \biggl(\sum_{k = 1}^{m} f\bigl(k\theta\bigr)\biggr)
                                                        g(m\theta)
     + \sum_{n = 1}^{[x/\theta]} \theta^2 \biggl(\sum_{k = 1}^{n} g\bigl(k\theta\bigr)\biggr)
                                                                             f(n\theta) \nonumber
     - \sum_{n = 1}^{[x/\theta]} \theta^2 f(n\theta)g(n\theta)\\
    &= R_\theta\bigl(R_\theta(f)g\bigr)(x) + R_\theta\bigl(fR_\theta(g)\bigr)(x) + \theta R_\theta(fg)(x).
\label{Riemsum1}
\end{align}}
Similarly for the map $R'_\theta$ except that the diagonal, counted twice, must be subtracted instead of added. Hence, the Riemann summation maps $R_\theta$ and $R'_\theta$ satisfy the weight $\theta$ and the weight $-\theta$ Rota--Baxter relation, respectively.

\begin{proposition}
Let $(A,R)$ be a Rota--Baxter algebra. The map $\tilde{R}=-\theta id_A-R$ is a Rota--Baxter 
map of weight $\theta$ on $A$. The images of $R$ and $\tilde{R}$, $A_{\mp} \subseteq A$, 
respectively are subalgebras in $A$.
\end{proposition}

We omit the proof since it follows directly from the Rota--Baxter relation. A {\textit{Rota--Baxter ideal}} of a Rota--Baxter algebra $(A,R)$ is an ideal $I \subset A$ such that $R(I)
\subseteq I$.

The Rota--Baxter relation extends naturally to the Lie algebra $L_A$ corresponding to $A$:
$$
    [R(x), R(y)] = R\bigl([R(x),y] + [x,R(y)]\bigr) + \theta R\bigl([x, y]\bigr)
$$
making $(L_A,R)$ into a Rota--Baxter Lie algebra of weight $\theta$. 

\begin{proposition}\label{prop:RBdouble}
The vector space underlying $A$ equipped with the product:
\begin{align}
        x \ast_\theta y := R(x)y + xR(y) + \theta xy
\label{def:RBdouble}
\end{align}
is again a Rota--Baxter algebra of weight $\theta$ with Rota--Baxter map $R$. We denote it by $(A_\theta,R)$ and call it {\textit{double Rota--Baxter algebra}}. The Rota--Baxter map $R$ becomes a (not necessarily unital even if $A$ is unital) algebra homomorphism from the algebra $A_{\theta}$ to $A$. 
\end{proposition}

Let us remark that for the corresponding Lie algebra $L_A$ we find the new Lie bracket (compare with \cite{STS}):
\begin{align*}
        [x, y]_\theta := [R(x),y] + [x,R(y)] + \theta [x,y].
\end{align*}
One sees immediately that 
$x \ast_\theta y = -\theta^{-1}\bigl(R(x)R(y)-\tilde{R}(x)\tilde{R}(y)\bigr)$, and
\begin{equation}
    R\bigl(a \ast_\theta b\bigr) = R(a)R(b)
    \;\;\ {\rm{and}}\ \;\;
    \tilde{R}\bigl( a \ast_\theta b \bigr) = -\tilde{R}(a)\tilde{R}(b).
  \label{eq:RBhom}
\end{equation}

The result in Proposition~\ref{prop:RBdouble} is best understood in the dendriform setting which we introduce now.  A {\sl dendriform algebra\/}~\cite{L} over a field $k$ is a $k$-vector space $A$ endowed with two bilinear operations $\prec$ and $\succ$ subject to the three axioms below:
 \allowdisplaybreaks{
\begin{align*}
 (a\prec b)\prec c = a\prec(b*c),      \quad
 (a\succ b)\prec c = a\succ(b\prec c), \quad
  a\succ(b\succ c) = (a*b)\succ c,
\end{align*}}
where $a*b$ stands for $a\prec b+a\succ b$. These axioms easily yield associativity for the law $*$. The bilinear
operations $\rhd$ and $\lhd$ defined by:
\begin{equation}
\label{def:prelie}
    a \rhd b:= a\succ b-b\prec a,
    \hskip 12mm
    a \lhd b:= a\prec b-b\succ a
\end{equation}
are left pre-Lie and right pre-Lie, respectively, which means that we have:
 \allowdisplaybreaks{
\begin{align}
    (a\rhd b)\rhd c-a\rhd(b\rhd c)&= (b\rhd a)\rhd c-b\rhd(a\rhd c),\label{prelie}\\
    (a\lhd b)\lhd c-a\lhd(b\lhd c)&= (a\lhd c)\lhd b-a\lhd(c\lhd b).
\end{align}}
The associative operation $*$ and the pre-Lie operations $\rhd$, $\lhd$ all define the same Lie bracket:
\begin{equation}
    [a,b]:=a*b-b*a=a\rhd b-b\rhd a=a\lhd b-b\lhd a.
\end{equation}

\begin{proposition}\label{prop:RBdend} \cite{E}
Any Rota--Baxter algebra gives rise to two dendriform algebra structures given by:
 \allowdisplaybreaks{
\begin{align}
    a\prec b &:= aR(b)+\theta ab=-a\tilde{R}(b),\hskip 8mm a\succ b:=R(a)b,   \label{RBpreLie1}\\
    a\prec' b&:= aR(b),\hskip 33mm a\succ' b:=R(a)b+\theta ab=-\tilde{R}(a)b. \label{RBpreLie2}
\end{align}}
\end{proposition}
The associated associative product $*$ is given for both structures by $a*b=aR(b)+R(a)b+\theta ab$ and thus coincides with the double Rota--Baxter product~(\ref{def:RBdouble}). 

\begin{remark}\cite {E}
In fact, by splitting again the binary operation $\prec$ (or alternatively $\succ'$), any Rota--Baxter algebra is {\rm tri-dendriform\/}~\cite{LR2}, in the sense that the Rota--Baxter structure yields three binary operations $<,\diamond$ and $>$ subject to axioms refining the axioms of dendriform algebras. The three binary operations are defined by $a < b = aR(b)$, $a \diamond b=\theta ab$ and $a > b = R(a)b$. Choosing to put the operation $\diamond$ to the $<$ or $>$ side gives rise to the two dendriform structures above.

\end{remark}
Let $\overline A=A\oplus k.\un$ be our dendriform algebra augmented by a unit $\un$:
\begin{equation}
\label{unit-dend}
    a \prec \un := a =: \un \succ a
    \hskip 12mm
    \un \prec a := 0 =: a \succ \un,
\end{equation}
implying $a*\un=\un*a=a$. Note that $\un*\un=\un$, but that $\un\prec \un$ and $\un\succ \un$ are not defined \cite{R}, \cite{C}. We recursively define the following set of elements of $\overline A[[t]]$ for a fixed $x\in A$:
 \allowdisplaybreaks{
\begin{align*}
    w^{(0)}_{\prec}(x)  &= w^{(0)}_{\succ}(x)=\un, \\
    w^{(n)}_{\prec}(x) &:= x \prec \bigl(w^{(n-1)}_\prec(x) \bigr), \\
    w^{(n)}_{\succ}(x) &:= \bigl(w^{(n-1)}_\succ(x)\bigr)\succ x.
\end{align*}}
We also define the following set of iterated left and right pre-Lie products~(\ref{def:prelie}). For $n>0$, let $a_1,\ldots ,a_n \in A$:
 \allowdisplaybreaks{
\begin{align}
    \ell^{(n)}(a_1,\dots,a_n) &:=
    \Bigl( \cdots \bigl( (a_1 \rhd a_2) \rhd a_3 \bigr)
    \cdots \rhd a_{n-1} \Bigr) \rhd a_n
\label{leftRBpreLie}\\
    r^{(n)}(a_1,\dots,a_n) &:=
    a_1 \lhd \Bigl( a_2 \lhd \bigl( a_3 \lhd
    \cdots (a_{n-1} \lhd a_n) \bigr) \cdots \Bigr).
\label{rightRBpreLie}
\end{align}}
For a fixed single element $a \in A$ we can write more compactly for $n>0$:
 \allowdisplaybreaks{
\begin{align}
    \ell^{(n+1)}(a) = \bigl(\ell^{(n)}(a)\bigr)\rhd a
		\quad\ {\rm{ and }} \quad\
    r^{(n+1)}(a) = a \lhd \bigl(r^{(n)}(a)\bigr)
\label{def:pre-LieWords}
\end{align}}
and $\ell^{(1)}(a):=a=:r^{(1)}(a)$. We have the following theorem
\cite{EMP07b, EGP}.

\begin{theorem} \label{thm:ncSpitzerSpecial} We have:
 \allowdisplaybreaks{
\begin{align*}
    w^{(n)}_{\succ}(a) &=
    \sum\limits_{i_1+ \cdots + i_k=n \atop  i_1,\ldots,i_k>0}
    \frac{\ell^{(i_1)}(a)
    * \cdots *
    \ell^{(i_k)}(a)}
         {i_1(i_1+i_2)\cdots(i_1+\cdots+i_k)},
         \\
    w^{(n)}_{\prec}(a) &=
    \sum\limits_{i_1+ \cdots +i_k=n \atop  i_1,\ldots,i_k>0}
    \frac{r^{(i_k)}(a)
    * \cdots *
    r^{(i_1)}(a)}
         {i_1(i_1+i_2)\cdots(i_1+\cdots+i_k)}.
\end{align*}}
\end{theorem}

These identities nicely show how the dendriform pre-Lie and associative products fit together. This will become even more evident in the following.  

We are interested in the solutions $X$ and $Y$ in $\overline A[[t]]$ of the following two equations:
\begin{equation}
\label{eq:prelie}
     X = \un + ta \prec X,
    \hskip 12mm
     Y = \un - Y \succ ta.
\end{equation}
Formal solutions to (\ref{eq:prelie}) are given by:
\begin{equation*}
    X = \sum_{n \geq 0} t^nw^{(n)}_{\prec}(a)
    \hskip 15mm {\rm{resp.}} \hskip 15mm
    Y = \sum_{n \geq 0} (-t)^nw^{(n)}_{\succ}(a).
\end{equation*}
Let us introduce the following operators in $A$, where $a$ is any
element of $A$:
 \allowdisplaybreaks{
\begin{align*}
    L_\prec[a](b)&:= a \prec b \hskip 4mm L_\succ[a](b):= a\succ b \hskip 4mm
    R_\prec[a](b):= b \prec a \hskip 4mm R_\succ[a](b):= b\succ a\\
    L_\lhd[a](b)&:= a \lhd b \hskip 4mm L_\rhd[a](b) := a \rhd b \hskip 4mm
    R_\lhd[a](b):=b \lhd a \hskip 4mm R_\rhd[a](b):= b \rhd a.
\end{align*}}

We have recently obtained the following {\sl pre-Lie Magnus expansion\/} \cite{EM}:

\begin{theorem} \label{thm:main}
Let $\Omega':=\Omega'(ta)$, $a \in A$, be the element of $tA[[t]]$ such that $X=\exp^*(\Omega')$ and $Y=\exp^*(-\Omega')$, where $X$ and $Y$ are the solutions of the two equations (\ref{eq:prelie}), respectively. This element obeys the following recursive equation:
 \allowdisplaybreaks{
\begin{align}
    \Omega'(ta) &= \frac{R_\lhd[\Omega']}{1-\exp(-R_\lhd[\Omega'])}(ta)=
                \sum_{m \ge 0}(-1)^m \frac{B_m}{m!}R_\lhd[\Omega']^m(ta),\label{main1}
%\\
%    \Omega'(ta)&=& \frac{L_\rhd[\Omega']}{1-\exp(-L_\rhd[\Omega'])}(at)=
%                \sum_{n\ge 0}(-1)^n\frac{B_n}{n!}L_\rhd[\Omega']^n(at),\label{main2}
\end{align}}
or alternatively:
 \allowdisplaybreaks{
\begin{align}
    \Omega'(ta) &= \frac{L_\rhd[\Omega']}{\exp(L_\rhd[\Omega'])-1}(ta)
            =\sum_{m\ge 0}\frac{B_m}{m!}L_\rhd[\Omega']^m(ta),    \label{main3}
%\\
%    \Omega'(at)&=& \frac{R_\lhd[\Omega']}{\exp(R_\lhd[\Omega'])-1}(at)
%            =\sum_{n\ge 0}\frac{B_n}{n!}R_\lhd[\Omega']^n(at),\label{main4}
\end{align}}
where the $B_l$'s are the Bernoulli numbers.
\end{theorem}

Recall that the Bernoulli numbers are defined via the generating series:
$$
    \frac{z}{\exp(z)-1} = \sum_{m \ge 0} \frac{B_m}{m!}z^m
                        = 1 - \frac{1}{2}z + \frac{1}{12}z^2 - \frac{1}{720}z^4 + \cdots,
$$
and observe that $B_{2m+3}=0$, $m \geq 0$.

Suppose that the dendriform structure of $A$ comes from a unital Rota--Baxter algebra of weight $\theta$. The unit (which we denote by $1$) has nothing to do with the artificially added unit $\un$ of the underlying dendriform algebra. We extend the Rota--Baxter algebra structure to $\overline A$ by setting:
\begin{equation*}
	R(\un):=1,
	 \hskip 8mm 
	\widetilde R(\un):=-1 
	 \hskip 5mm \hbox{ and }
	\un.x = x.\un =0 
	 \hbox{ for any } 
	x\in\overline A.
\end{equation*}
This is consistent with the axioms (\ref{unit-dend}) which in particular yield $\un \succ x=R(\un)x$ and $x\prec \un=-x \widetilde R(\un)$, in coherence with the dendriform axioms. Using (\ref{RBpreLie1}) the pre-Lie products~(\ref{def:prelie}) write:
 \allowdisplaybreaks{
\begin{align}
    a \rhd_\theta b &= a\succ b-b\prec a = R(a)b + b\tilde{R}(a) = [R(a),b] - \theta ba 
\label{RBpre-Lie1}\\
    a \lhd_\theta b &= a\prec b-b\succ a = -a\tilde{R}(b) - R(b)a = [a,R(b)] + \theta ab.
\label{RBpre-Lie2}
\end{align}}

Introduce the {\sl{weight $\theta \in k$ pre-Lie Magnus type recursion}}, $\Omega'_\theta:=\Omega'_\theta(ta) \in tA[[t]]$, where the Rota--Baxter operator $R$ is naturally extended to $\overline A[[t]]$ by $k[[t]]$-linearity:
 \allowdisplaybreaks{
\begin{align}
    \Omega'_\theta(ta) &= \sum_{m \ge 0} \frac{B_m}{m!}L_{\rhd_\theta}[\Omega'_\theta]^m(ta).
\label{RB-MagunsMain}
\end{align}}

First, we observe that the weight zero case, $\theta=0$, is fully coherent with the origin of Magnus' work~\cite{Magnus}.
 
\begin{corollary}\label{cor:Magnus-Zero}
The limit $\theta \to 0$ of the pre-Lie Magnus type recursion (\ref{RB-MagunsMain}) reduces to the classical Magnus expansion:
\begin{equation}
\label{MagnusClassic}
	    \Omega'_0(ta) = \sum_{m \ge 0} \frac{B_m}{m!} ad^{(m)}_{R(\Omega'_0)}(ta).
\end{equation}
\end{corollary}

Here, as usual, $ad_f(g):=fg-gf:=[f,g]$. This follows immediately from (\ref{RBpre-Lie1}), i.e. $a \rhd_0 b=[R(a),b]$. Recall that Magnus in~\cite{Magnus} considers solutions of the classical initial value problem:
\begin{equation}
\label{IVP1}
	\frac{d}{ds}\Phi(s) = \Psi(s)\Phi(s),\ \qquad\ \ \Phi(0)=1
\end{equation} 
in a non-commutative context, i.e. $\Psi$ and $\Phi$ are supposed to be linear operators depending on a real variable $t$. Here, $1$ denotes the identity operator. Magnus obtained a differential equation for the linear operator $\Omega(\Psi)(s)$ depending on $\Psi(s)$, and with $\Omega(0)=0$, such that:
$$
    Y(s) = \exp\bigl(\Omega(\Psi)(s)\bigr) 
         = \exp\Bigl( \int^s_0 \dot{\Omega}(\Psi)(u)\, du \Bigr)
         = \sum_{n \ge 0} \frac{\bigl(\Omega(\Psi)(s)\bigr)^n}{n!},
$$
leading to the recursively defined classical Magnus expansion:
\begin{equation}
\label{MagnusOmega}
    \Omega(t\Psi)(s) = t\int_0^s \Psi(u)\, du 
    + \int^s_0 \sum_{n > 0} \frac{B_n}{n!} ad^{(n)}_{\Omega(t\Psi)(u)}(t\Psi(u))\, du,
\end{equation}
where we introduced the parameter $t$ for later use. The expansion~(\ref{MagnusClassic}}) coincides with (\ref{MagnusOmega}) if the underlying weight zero Rota--Baxter algebra is the one mentioned above, that is, the ring of real continuous functions $C(\mathbb{R})$ with pointwise product and the indefinite Riemann integral as weight zero Rota--Baxter map. Sometimes, Magnus' expansion (\ref{MagnusOmega}) is also called {\it{continuous Baker--Campbell--Hausdorff formula}}, e.g. see~\cite{gelfand1995,miel1970,strich1987}.

\medskip

Using (\ref{eq:RBhom}), Theorem~\ref{thm:main} implies for a fixed element $a \in A$ the following corollary which can be interpreted as the non-commutative Spitzer identity.

\begin{corollary}\label{cor:RBrecursions}
Let $(A,R)$ be a Rota--Baxter algebra of weight $\theta$. The elements $\hat{X}:=-\widetilde R(X)=\exp(-\tilde{R}(\Omega'_\theta(ta)))$ and
$\hat{Y}:=R(Y)=\exp(-R(\Omega'_\theta(ta)))$ in $A[[t]]$ solve the
equations:
\begin{equation}
\label{RBrecursions}
    \hat{X} = 1 - t\tilde{R}(a\hat{X}) 
    \hskip 15mm {\rm{resp.}} \hskip 15mm 
    \hat{Y} = 1 - tR(\hat{Y}a).
\end{equation}
\end{corollary}

Moreover, these recursions lead to the following theorem due to Atkinson~\cite{A}.

\begin{theorem}\label{thm:Atkinson}
Let $(A,R)$ be a Rota--Baxter algebra of weight $\theta$. The recursions (\ref{RBrecursions}) 
have the factorization property:
$$
	\hat{Y}(1-ta\theta)\hat{X}=1 \;\; {\rm{or}}\;\; (1-ta\theta)= \hat{Y}^{-1}\hat{X}^{-1}.
$$
\end{theorem}

\begin{proof}  Here, $1$ is the algebra unit in $A$. The proof of this statement reduces to a simple algebraic exercise. Recall that $\tilde{R}=-\theta id_A -R$, then:
 \allowdisplaybreaks{
\begin{align*}
	\hat{Y}\hat{X} &= \bigl(1 - tR(\hat{Y}a)\bigr)\bigl(1 - t\tilde{R}(a\hat{X})\bigr) \\
	               	     &= 1 - t\tilde{R}(a\hat{X}) - tR(\hat{Y}a) + t^2R(\hat{Y}a)\tilde{R}(a\hat{X})\\
			     &= 1 - t\tilde{R}\bigl((1 - tR(\hat{Y}a)) a \hat{X}\bigr) 
								       - tR\bigl(\hat{Y}a(1 - tR(a \hat{X}))\bigr) = 1 + t\theta \hat{Y}a\hat{X}.       	
\end{align*}}
Uniquness of the factorization follows when $R$ is idempotent.  
\end{proof}

Without problems the reader verfies the next corollary.

\begin{corollary} \label{cor:invRBRec}
The elements $\hat{X}^{-1}=\exp(\tilde{R}(\Omega'_\theta(ta)))$ and
$\hat{Y}^{-1}:=R(Y)=\exp(R(\Omega'_\theta(ta)))$ in $A[[t]]$ solve the
equations:
\begin{equation}
\label{RBrecursionsInverse}
    \hat{X}^{-1} = 1 + t\tilde{R}(\hat{Y}a) 
    \hskip 15mm {\rm{resp.}} \hskip 15mm 
    \hat{Y}^{-1} = 1 + tR(a\hat{X}).
\end{equation}
\end{corollary}

Hence, defining the application $\bar{B}(a):=\hat{Y}a$ we may state the two key equations:
$$
   \hat{Y} = 1 - tR(\bar{B}(a)) 
    \;\; {\rm{and}}\ \;\; 
   \hat{X}^{-1} = 1 + t\tilde{R}(\bar{B}(a)).
$$ 
Corollary~\ref{cor:RBrecursions} immedialtely results in the following lemma.

\begin{lemma}\label{lem:RBBogoPrep}
\begin{equation}
\label{RBBogoPrep}
	\bar{B}(ta)=\hat{Y}ta=\exp^{*_\theta}(\Omega'_\theta(ta))-1.
\end{equation}
\end{lemma}
\begin{proof} Indeed, using simple algebra we see that: 
 \allowdisplaybreaks{
\begin{align*}
	\exp^{*_\theta}(\Omega'_\theta(ta))-1 &=  \sum_{n=1}^{\infty} \frac{\bigl(\Omega'_\theta(ta)\bigr)^{{*_\theta}n}}{n!}\\
								&= \sum_{n=1}^{\infty} 
								     \frac{-\theta^{-1} }{n!}\Bigl( \bigl(R(\Omega'_\theta(ta))\bigr)^n 
								      -(-1)^n \bigl(\tilde{R}(\Omega'_\theta(ta))\bigr)^n\Bigr)\\
								&= - \sum_{n=0}^{\infty} 
									\frac{\theta^{-1}}{ n!}\bigl(R(\Omega'_\theta(ta)\bigr)^n 
									+ \sum_{n=0}^{\infty} \frac{(-1)^{n}\theta^{-1}}{ n!}
									    \bigl(\tilde{R}(\Omega'_\theta(ta)\bigr)^n\\
								&= -\exp\bigl(R(\Omega'_\theta(ta))\bigr)	 
								     +\exp\bigl(-\tilde{R}(\Omega'_\theta(ta))\bigr)\\
								&= \exp\bigl(R(\Omega'_\theta(ta))\bigr) \bigl(-1 +    
											     \exp\bigl(-R(\Omega'_\theta(ta))\bigr)
											     \exp\bigl(-\tilde{R}(\Omega'_\theta(ta))\bigr)\Bigr)	\\
								&= \exp\bigl(R(\Omega'_\theta(ta))\bigr)ta = t\hat{Y}a. 
\end{align*}} 
\end{proof}

Let us take a closer look at Atkinson's theorem. In the light of Corollary~\ref{cor:RBrecursions} 
we find by using the exponential solutions $\hat{X}:=-\widetilde R(X)=\exp(-\tilde{R}(\Omega_\theta(ta)))$ 
and $\hat{Y}:=R(Y)=\exp(-R(\Omega_\theta(ta)))$ in $A[[t]]$ to the recursions~(\ref{RBrecursions}):
\begin{equation}
\label{ncSpitzerMagnus}
    1-\theta t a = \exp(-\theta\alpha_\theta)
                 = \exp\Bigl(R\bigl(\Omega'_\theta(ta)\bigr)\Bigr)
  	               \exp\Bigl(\tilde{R}\bigl(\Omega'_\theta(ta)\bigr)\Bigr)
\end{equation}
with $\alpha_\theta:=\alpha_\theta(t a):=-\frac{1}{\theta}\log (1 - \theta t a)$. 

When $\theta=0$ we see immediately that $R=-\tilde{R}$ and the above factorization of the algebra unit becomes very evident. Instead, let us keep $\theta \neq 0$, but assume the underlying Rota--Baxter algebra to be commutative. Recall that $a \rhd_\theta b=  - \theta ba$. Hence, we see that:
$$
    \Omega'_\theta(ta) = \sum_{m \ge 0} \frac{B_m}{m!}\bigl(-\theta \Omega'_{\theta}\bigr)^m ta.
$$ 
Which, using the generating series for the Bernoulli numbers, is solved by $\Omega'_\theta(ta)=-\theta^{-1}\log(1-\theta ta)$. Hence, in the commutative setting we find: 
$$
	1-\theta ta= \exp\Bigl(R\bigl(-\theta^{-1}\log(1-\theta ta)\bigr)\Bigr)
	               \exp\Bigl(\tilde{R}\bigl(-\theta^{-1}\log(1-\theta ta)\bigr)\Bigr).	
$$
which is in full accordance with the classical result due to Spitzer~\cite{Spitzer}. In fact, Baxter~\cite{B} showed for commutative Rota--Baxter algebras $(A,R)$ of weight $\theta$ the identity:  
$$
    \exp\Big(\sum_{n>0}\frac{r_n t^n}{n}\Big) = 1 + \sum_{m > 0} a_m t^m
$$
in $A[[t]]$, where $r_n:=R(\theta^{n-1}a^n)$, $a_1 = r_1 = R(a)$, and:
$$
    a_m = \sum_{(\lambda_1,\dots,\lambda_m)}
    \frac{r_1^{\lambda_1} \cdots r_m^{\lambda_m}}{1^{\lambda_1}2^{\lambda_2}
    \cdots m^{\lambda_m} \lambda_1! \cdots \lambda_m!}
    = \underbrace{R\bigl(R(R( \cdots R}_{m-times}(a)a)\dots a)a\bigr).
$$
The sum goes over all integer $m$-tuples $(\lambda_1,\dots,\lambda_m)$, $\lambda_i \geq 0$ for which $1 \lambda_1 + \dots + m \lambda_m = m$. The non-commutative generalization of this formulation of Spitzer's identity follows from Theorem~\ref{thm:ncSpitzerSpecial} together with Proposition~\ref{prop:RBdend}. Indeed, with $\Omega'(ta)=\sum_{n>0} t^n \Omega'_{(n)}(a)$ we find:
 \allowdisplaybreaks{
\begin{align*}
    \exp\Big(-\sum_{n>0} R(\Omega'_{(n)}) t^n\Big) &= 1 + \sum_{m > 0} R(w^{(m)}_{\succ}(a)) (-t)^m\\	
    \exp\Big(-\sum_{n>0} \tilde{R}(\Omega'_{(n)}) t^n\Big) &= 1 + \sum_{m > 0} \tilde{R}(w^{(m)}_{\prec}(a)) (-t)^m		 
\end{align*}
corresponding to the recursions:
$$
	\hat{X} = 1 - R(\hat{X}a) \;\; {\rm{ resp. }}\;\; \hat{Y} = 1 - \tilde{R}(a\hat{Y}).
$$
Again, but more concretely, in the non-commutative case we see how both, the double Rota--Baxter product (\ref{def:RBdouble}) and the pre-Lie product (\ref{RBpre-Lie1}) fit together in the algebraic structure 
of weight $\theta$ Rota--Baxter algebras.

\subsection{Weight $\theta$ BCH-recursion vs. pre-Lie Magnus expansion}

In the light of Corollary~\ref{cor:birkhoffBCH} and Atkinson's Theorem, it seems to be natural to compare the weight 
$\theta$ pre-Lie Magnus expansion with the Baker--Campbell--Hausdorff recursion~(\ref{BCHrecursion2}). 

For this, we first introduce the weight $\theta$ Baker--Campbell--Hausdorff recursion following simply from a linear fitration preserving map $P$, such that $-\theta id_A = P + (-\theta id_A - P)$, $\theta \in k$. Hence:
\begin{equation}
    \label{thetaBCHrecursion}
    \chi_\theta(a) = a - \frac{1}{\theta}\mop{BCH}\bigl(-P(\chi_\theta(a)),\ -\theta a\bigr).
\end{equation}
Giving rise to the factorization:
\begin{equation}
\label{birkhoffTheta}
	\exp(-\theta a)=\exp\Bigl(P\bigl(\chi_\theta(a)\bigr)\Bigr)
	                \exp\Bigl(\tilde{P}\bigl(\chi_\theta(a)\bigr)\Bigr).
\end{equation}
Now, let us assume that $P$ is a filtration preserving Rota--Baxter map. Then Corollary~\ref{cor:birkhoffBCH} and Theorem~\ref{thm:Atkinson}, respectively equation (\ref{ncSpitzerMagnus}) imply the equality:
\begin{equation}
\label{chiVSMagnus}
    \Omega'_\theta(t a)
           = \chi_{\theta}(\alpha_\theta)
           = \chi_\theta\Bigl(-\frac{\log (1-\theta t a)}{\theta}\Bigr),
\end{equation}
From~(\ref{chiVSMagnus}) we get for any $\alpha \in t A[[\lambda]]$:
\begin{equation}
\label{chiVSMagnusbis}
    \chi_\theta\big(\alpha_\theta) = \Omega'_\theta\Bigl(\frac{1-\exp(-\theta\alpha)}{\theta}\Bigr).
\end{equation}

\subsection{Application to perturbative renormalisation II}
\label{subsect: appl II}

We return to paragraph \ref{subsect: appl I} where we analysed Connes--Kreimer's factorization from the point of view of the Baker--Campbell--Hausdorff recursion (\ref{BCHrecursion2}). Recall that the projector $\pi$ (respectively its lift to $\mathcal L=\mathcal L(\mathcal H,\mathcal A)$, denoted by $P$) is a weight minus one Rota--Baxter map. Hence the Birkhoff--Connes--Kreimer factorization naturally fits into the context of Rota--Baxter algebra, in particular with respect to Atkinson's factorization theorem respectively the non-commutative Spitzer identity. Hence, it follows ultimately that the group $G_1(\mathcal A)$ decomposes as a set into the product of two subgroups:
$$
	G_1(\mathcal A) = G_1^-(\mathcal A) \ast G_1^+(\mathcal A), 
	\;\; {\rm{where}}\ \;\;\ 
	G_1^-(\mathcal A) = \exp^*\bigl(P(\mathcal A)\bigr), \;\;\; G_1^+(\mathcal A) =\exp^*\bigl(\tilde{P}(\mathcal A)\bigr).
$$

Now, using (\ref{chiVSMagnusbis}) we see that the counterterm character $\varphi_-$ in the decomposition:
$$
	\varphi = \exp^*(a) = \varphi_-^{-1} * \varphi_+  
$$
writes:
$$
	\varphi_- = \exp^*\Bigl(-P\bigl(\Omega'_{-1}(\exp^*(a)-e)\bigr)\Bigr)
$$ 
where $ \exp^*(a)-e = \varphi - e$ already appeared in the context of (\ref{pre-spitzer-}) respectively (\ref{pre-spitzer+}), see also Proposition~\ref{prop:exp}. Using Lemma~\ref{lem:RBBogoPrep} Bogoliubov's preparation map $B(\varphi):=\varphi_- * (\varphi - e)$ finds its exponential form:
$$
	B(\varphi)= \exp^{*_{-1}}\bigl(\Omega'_{-1}( \varphi -e )\bigl)-e.
$$  
Such that the non-commutative Spitzer identity implies Bogoliubov's recursion for $\varphi_-$. Moreover,  Corollary~\ref{cor:invRBRec} tells immediately the equation for $\varphi_+$. Here, ${*_{-1}}$ stands for the Rota--Baxter double product in the weight minus one Rota--Baxter algebra $(\mathcal L(\mathcal H,\mathcal A),P)$. 

Recall our characterization of the Baker--Campbell--Hausdorff recursion (\ref{BCHrecursion2}) as the analog of Bogoliubov's preparation map on the Lie algebra $\mathfrak g_1(\mathcal A)$ of infinitesimal characters. Now, by recalling Proposition~\ref{prop:exp} we may identify the weight minus one pre-Lie Magnus expansion $\Omega'_{-1}$ as the analog of Bogoliubov's preparation map. More precisely, first remember that $\mathfrak g(\mathcal A)$ contains $\mathfrak g_1(\mathcal A)$ as a Lie subalgebra. Hence, let $\varphi \in G_1(\mathcal A) \subset G(\mathcal A)$, i.e. $\varphi = e + ( \varphi-e)$, where obviously $a:=\varphi - e \in \mathfrak g(\mathcal A)$. Then:
\begin{align}
			 \Omega'_{-1} (a)  = \sum_{m \ge 0} \frac{B_m}{m!}L_{\rhd_{-1}}[\Omega'_{-1}]^m(a),
\label{LieBogoliubov}
\end{align}
maps $a \in \mathfrak g(\mathcal A) \to \mathfrak g_1(\mathcal A)$, such that $B(\varphi)= \exp^{*_{-1}}\bigl(\Omega'_{-1}( \varphi-e )\bigr)-e$ and 
$$
	\varphi_-=\exp^*\Bigl( - P\bigl(\Omega'_{-1}( \varphi-e )\bigr) \Bigr)
	\;\; {\rm{and}} \;\; 
	\varphi_+=\exp^*\Bigl( \tilde{P}\bigl(\Omega'_{-1}( \varphi-e )\bigr) \Bigr)
$$
solve: 
$$
	\varphi_- = e - P\bigl(\varphi_- * (\varphi-e)\bigr)
	\;\; {\rm{and}} \;\; 
	\varphi_+ = e + \tilde{P}\bigl(\varphi_- * (\varphi-e)\bigr),
$$
respectively.

\smallskip

We remark here that this purely Lie algebraic approach to renormalization is an extension of earlier
work~\cite{EGP} and will be further explored in the near future. Let us mention that the results in \cite{EGP} rely on both, the properties of the Dynkin idempotent and on properties of
Hopf algebras encapsulated in the notion of associated descent algebras. Similarly, in~\cite{EMP}, see also \cite{EGP2}, we use free Lie algebra theory, i.e. Lie idempotents to achieve a closed form for the Bogoliubov recurison.

\subsection{Non-commutative Bohnenblust--Spitzer formulas}
\label{subsect:ncBSidentity}

Let $n$ be a positive integer, and let $\mathcal{OP}_n$ be the set of ordered partitions of $\{1,\ldots ,n\}$, i.e. sequences $(\pi_1,\ldots ,\pi_k)$ of disjoint subsets ({\sl blocks\/}) whose union is $\{1,\ldots ,n\}$. We denote by  $\mathcal{OP}_n^k$ the set of ordered partitions of $\{1,\ldots ,n\}$ with $k$ blocks. Let us introduce for any $\pi\in\mathcal{OP}_n^k$ the coefficient:
\begin{equation*}
	\omega(\pi)=\frac{1}{|\pi_1|(|\pi_1|+|\pi_2|)\cdots (|\pi_1|+|\pi_2|+\cdots+|\pi_k|)}.
\end{equation*}

\begin{theorem}
Let $a_1,\ldots ,a_n$ be elements in a dendriform algebra $A$. For any subset $E=\{j_1,\ldots, j_m\}$ of $\{1,\ldots ,n\}$ let $\mathfrak l(E)\in A$ defined by:
\begin{equation*}
	\mathfrak l(E):=\sum_{\sigma\in S_m}\mathfrak l^{(m)}(a_{j_{\sigma_{1}}},\ldots,a_{j_{\sigma_{m}}}).
\end{equation*}
we have:
\begin{equation*}
	\sum_{\sigma\in S_n}\Bigl (\ldots (a_{\sigma_1}\succ a_{\sigma_2})\succ \cdots
	                                                      a_{\sigma_{n-1}}\Big)\succ a_{\sigma_n}
	=\sum_{k\ge 1} \sum_{\pi\in\mathcal{OP}_n^k}\omega(\pi)\mathfrak l(\pi_1)* \cdots *\mathfrak l(\pi_k).
\end{equation*}
\end{theorem}

See \cite{EMP} where this identity is settled in the Rota--Baxter setting, see also~\cite{EGP2}. The proof in the dendriform context is entirely similar. Another expression for the left-hand side can be obtained \cite{EMP07b}: For
any permutation $\sigma\in S_n$ we define the element $T_\sigma(a_1,\ldots ,a_n)$ as follows: define first the subset $E_\sigma\subset\{1,\ldots ,n\}$ by $k\in E_\sigma$ if and only if
$\sigma_{k+1} > \sigma_j$ for any $j\le k$. We write $E_\sigma$ in
the increasing order:
\begin{equation*}
    1\le k_1<\cdots < k_p\le n-1.
\end{equation*}
Then we set:
\begin{equation}
\label{tsigma}
 T_\sigma(a_1,\ldots ,a_n):=
   \ell^{(k_1)}(a_{\sigma_1},\dots,a_{\sigma_{k_1}}) *\cdots *
   \ell^{(n-k_{p})}(a_{\sigma_{k_p+1}},\dots,a_{\sigma_{n}})
%
%\Big(...\big((a_{\sigma_1}\rhd a_{\sigma_2})\rhd\cdots\big)\rhd
%a_{\sigma_{k_1}}\Big)*\cdots *\Big(...\big((a_{\sigma_{k_p+1}}\rhd
%a_{\sigma_{k_p+2}})\rhd \cdots\big)\rhd a_{\sigma_{n}}\Big).
\end{equation}
There are $p+1$ packets separated by $p$ stars in the right-hand side of the expression (\ref{tsigma}) above, and the parentheses are set to the left inside each packet. Following \cite{Lam} it is convenient to write a permutation by putting a vertical bar after each element of $E_\sigma$. For example for the permutation $\sigma=(3261457)$ inside $S_7$ we have $E_\sigma=\{2,6\}$. Putting the vertical bars:
\begin{equation*}
	\sigma=(32|6145|7)
\end{equation*}
we see that the corresponding element in $A$ will then be:
 \allowdisplaybreaks{
\begin{align*}
 T_\sigma(a_1,\ldots , a_7)&=\ell^{(2)}(a_3,a_2) * \ell^{(4)}(a_6,a_1,a_4,a_5)* \ell^{(1)}(a_7)\\
                           			&= (a_3\rhd a_2)*\Big(\big((a_6\rhd a_1)\rhd a_4\big)\rhd a_5\Big)*a_7.
\end{align*}}

\begin{theorem}\label{main}
For any $a_1,\ldots ,a_n$ in the dendriform algebra $A$ the following identity holds:
 \allowdisplaybreaks{
\begin{align}
 \sum_{\sigma\in S_n}\big(\cdots(a_{\sigma_1}\succ a_{\sigma_2})\succ\cdots\big)\succ a_{\sigma_n}
    &=\sum_{\sigma\in S_n}T_\sigma(a_1,\ldots ,a_n).
\label{eq:main}
\end{align}}
\end{theorem}

A $q$-analog of this identity has been recently proved by J-C. Novelli and J-Y. Thibon \cite{NT}.

\section{As simple as it gets: a matrix calculus for renormalization}
\label{sect:matrix}

We shortly introduce a matrix setting for renormalization, associated with any left co-ideal of the Hopf algebra. Although we won't detail this point, let us mention that this matrix approach is particularly well-suited for the study of the renormalization group and the beta-function for local characters in
connected {\sl graded\/} Hopf algebras with values into meromorphic functions \cite{CK2}. See \cite{EM06} as well as \cite{EGGV,EG2} for a detailed account and applications.

\subsection{The matrix representation}
\label{subsect:matrix}

In this section we introduce the matrix representation of $\mathcal L(\mathcal H,\mathcal A)$ associated with a left coideal, following \cite{EM06}. Let $\mathcal H$ be a connected filtered Hopf algebra over the field $k$, let $\mathcal A$ be any commutative unital $k$-algebra, and let $\big(\mathcal L(\mathcal H,\mathcal A),\star\big)$ be the algebra of $k$-linear maps from $\mathcal H$ to $\mathcal A$ endowed with the convolution product. Let $J$ be any left coideal of $\mathcal H$ (i.e. a vector subspace of $\mathcal H$ such that $\Delta(J)\subset \mathcal H\otimes J$).

\medskip

We fix a basis $X=(x_i)_{i\in I}$ of the left coideal $J$.  Furthermore we suppose that this basis is denumerable (hence indexed by $I=\mathbb{N}$ or $I=\{1,\cdots ,m\})$ and {\sl filtration ordered\/}, i.e. such that if $i\le j$ and $x_j\in\mathcal H^{n}$, then $x_i\in\mathcal
H^{n}$.

\begin{definition} \label{def:coproductmatrix}
The {\sl coproduct matrix\/} in the basis $X$ is the $|I|\times |I|$ matrix $M$ with entries in $\mathcal H$ defined by~:
$$
    \Delta(x_i)=\sum_{j\in I}M_{ij}\otimes x_j.
$$
\end{definition}
The coproduct matrix is lower-triangular with diagonal terms equal to $\un$ (\cite{EM06} Lemma 1). Now define $\Psi_J:\mathcal L(\mathcal H,\mathcal A) \to \mop{End}_{\mathcal A}(\mathcal A\otimes J)$ by~:
\begin{equation}
\label{Psi}
    \Psi_J[f](x_j)=\sum_if(M_{ij})\otimes x_i.
\end{equation}
In other words, the matrix of $\Psi_J[f]$ is given by $f(M):=\big(f(M_{ij})\big)_{i,j\in I}$. It is shown in \cite{EG2} and also \cite{EM06} that the map $\Psi_J$ defined above is an algebra homomorphism. Its transpose does not depend on the choice of the basis. The Lie algebra of $\mathcal A$-valued infinitesimal characters (resp. the group of $\mathcal A$-valued characters) is mapped by $\Psi_J$ into the Lie subalgebra of strictly lower-triangular matrices (resp. into the group of lower triangular matrices
with --$\mathcal A$-algebra units-- $1$'s on the diagonal).

\medskip

The coproduct matrix $M$ with entries in $\mathcal H$ can be seen as the image of the identity map under  $\Psi_J:\mathcal L(\mathcal H,\mathcal H) \to \mop{End}_{\mathcal H}(\mathcal H \otimes J)$, i.e.~:
\begin{equation}
\label{Psi}
    \Psi_J[Id](x_j) = \sum_i Id(M_{ij})\otimes x_i.
\end{equation}
We have $\Psi_J[S]=M^{-1}$, where $S$ is the antipode. The matrix $L=\log M$ is the matrix of {\sl normal coordinates\/}. For any $\mathcal A$-valued character $\varphi$ we have~:
\begin{equation*}
	\log \Psi_J[\varphi]=\varphi(L).
\end{equation*}

\subsection{The matrix form of Connes--Kreimer's Birkhoff decomposition} 
\label{subsect:matrix}

Suppose that the commutative target space algebra $\mathcal A$ in
$\mathcal L(\mathcal H, \mathcal A)$ splits into two subalgebras:
\begin{equation}
\label{split}
    \mathcal A = \mathcal A_- \oplus \mathcal A_+,
\end{equation}
where the unit $1_{\mathcal A}$ belongs to $\mathcal A_+$. Let us denote by $\pi: \mathcal A \to \mathcal A_-$ the projection onto $\mathcal A_-$ parallel to $\mathcal A_+$, which is a weight $-1$ Rota--Baxter map. The algebra $\mathcal M:=\mathcal{M}^\ell_{|I|}(\mathcal A)$ of lower-triangular $|I| \times |I|$-matrices with coefficients in $\mathcal A$ is filtered by the subalgebras $\mathcal M^i=\{X\in\mathcal M,\,X_{kl}=0 \hbox{ if }k<l-i\}$. The filtration is finite, hence complete.

\smallskip

We define a Rota--Baxter map $\mathrm{R}$ on $\Psi_{J}[\mathcal L(\mathcal H,\mathcal A)] \subset \mathcal{M}^\ell_{|I|}(\mathcal A)$ by extending the Rota--Baxter map $\pi$ on $\mathcal A$ entrywise, i.e.,
for the matrix $\tau =(\tau_{ij}) \in \mathcal{M}^\ell_{|I|}(\mathcal A)$, define:
\begin{equation}
\label{RBonMatrices}
    \mathrm{R}(\tau) =  \big( \pi(\tau_{ij}) \big).
\end{equation}
The algebra $\mathcal M$ is then a complete filtered Rota--Baxter algebra. Let us denote $\Psi_J[\varphi]:=\widehat \varphi$ for short. As $\varphi\mapsto\widehat\varphi$ is a morphism of complete filtered Rota--Baxter algebras we immediately get the matrix Birkhoff--Connes--Kreimer
decomposition:
\begin{equation*}
	\widehat\varphi = \widehat{\varphi_-}^{-1}\widehat{\varphi_+}.
\end{equation*}
In other words, the map $\Psi_J$ respects the Birkhoff decomposition, i.e. $\widehat\varphi_\pm=\widehat{\varphi_\pm}$. We immediately see that $\widehat{\varphi}{}_-$ and $\widehat{\varphi}^{-1}_+$ are unique solutions of the following equations:
 \allowdisplaybreaks{
\begin{align}
    \widehat{\varphi}_-  &= {\bf{1}} - \mathrm{R}\Big(\widehat{\varphi}_- \ (\widehat{\varphi}-{\bf{1}})\Big),
    \label{eq:mphi-}\\
    \widehat{\varphi}^{-1}_+  &= {\bf{1}} - \tilde{\mathrm{R}}\Big( (\widehat{\varphi}-{\bf{1}})\ \widehat{\varphi}^{-1}_+ \Big),
    \label{eq:mphi+}
\end{align}}
respectively. Moreover, after some simple algebra using the matrix factorization 
$\widehat\varphi=\widehat\varphi_-^{-1}\widehat\varphi_{+}$~:
$$
  \widehat{\varphi}_+ (\widehat{\varphi}^{-1}-{\bf{1}})\ 
  = \widehat{\varphi}_{-} - \widehat{\varphi}_+ = - \widehat{\varphi}_{-}(\widehat{\varphi}-{\bf{1}})
$$
we immediately get the recursion for $\widehat{\varphi}_+$~\cite{EM06}:
\begin{equation}
\label{eq:mmphi+}
    \widehat{\varphi}_+  = 
    {\bf{1}} - \tilde{\mathrm{R}}\Big(\widehat{\varphi}_+ \ (\widehat{\varphi}^{-1}-{\bf{1}}) \Big),
\end{equation}
and hence we see that
\begin{equation}
\label{eq:mmmphi+}
    \widehat{\varphi}_+  
    = {\bf{1}} + \tilde{\mathrm{R}}\Big(\widehat{\varphi}_- \ (\widehat{\varphi}-{\bf{1}}) \Big).
\end{equation}

\noindent 
The matrix entries of $\widehat{\varphi}_-$ and $\widehat{\varphi}^{-1}_+$ can be calculated without recursions using $\sigma := \widehat{\varphi}$ from the equations~\cite{EGGV}:
 \allowdisplaybreaks{
\begin{align*}
    (\widehat{\varphi}_-)_{ij} &= -\pi(\sigma_{ij})\!
                                - \sum_{k=2}^{j-i}\: \sum_{i>l_1>l_2> \cdots >l_{k-1}>j} (-1)^{k+1}
       \pi\big(\pi(\cdots \pi(\sigma_{i l_{1}})\sigma_{l_{1} l_{2}}) \cdots \sigma_{ l_{k-1} j}\big)\\
    (\widehat{\varphi}^{-1}_+)_{ij} &= -\tilde{\pi}((\sigma^{-1})_{ij}) \\
      &  -\sum_{k=2}^{j-i}\: \sum_{i>l_1>l_2>\cdots > l_{k-1} > j}\!\! (-1)^{k+1}
        \tilde{\pi}\big(\tilde{\pi}(\cdots \tilde{\pi}
        ((\sigma^{-1})_{i l_{1}})(\sigma^{-1})_{l_{1} l_{2}})\cdots (\sigma^{-1})_{ l_{k-1} j}\big),
\end{align*}}
where $\tilde{\pi}:=id_\mathcal{A}-\pi$. The matrix entries of $\widehat{\varphi}_+$ follow from the first formula, i.e., the one for the entries in $\widehat{\varphi}_-$, by replacing $\pi$ by
$-\tilde{\pi}$. We may therefore define the matrix:
 \allowdisplaybreaks{
\begin{align} 
\label{BogoRbar}
 \widehat{B}[\varphi]:=\widehat{\varphi}_- \ (\widehat{\varphi}-{\bf{1}})
\end{align}}
such that:
 \allowdisplaybreaks{
\begin{align} \label{matrixAtkinson}
 \widehat{\varphi}_- = {\bf{1}} - \mathrm{R}\Big(\widehat B[ \varphi]\Big)\
 \makebox{ and }\
 \widehat{\varphi}_+ = {\bf{1}} + \tilde{\mathrm{R}}\Big(\widehat B[\varphi]\Big).
\end{align}}
In fact, equations (\ref{eq:mmmphi+}) and (\ref{eq:mphi+}) may be
called {\it{Bogoliubov's matrix formulae}\/} for the counter term
and renormalized Feynman rules matrix, $ \widehat{\varphi}_-$, $
\widehat{\varphi}_+$, respectively. Equation~(\ref{BogoRbar}) is
the matrix form of Bogoliubov's preparation
map~(\ref{bogo3}), e.g. see~\cite{C}~:
\begin{equation}
 \label{bogoliubov2}
    \widehat{B}[\varphi] := \Psi_{J}[B(\varphi)]
                                  = \Psi_{J}[\varphi_{-}\star(\varphi - e)].
\end{equation}

\begin{remark} \label{bch-CHI} {\rm{We may apply the result from subsection \ref{subsect:BCHrecursion} to the above matrix representation of $\mathfrak g_{\mathcal A}$ respectively $G_{\mathcal A}$. We have shown the existence of a unique non-linear map $\chi$ on $\mathfrak g_{\mathcal A}$ which allows to write the characters $\varphi_-$ and $\varphi_+$ as exponentials. In the
matrix picture we hence find for $\widehat{Z} \in  \widehat{\mathfrak g}_{\mathcal A}$ and  $\widehat{\varphi}=\exp(\widehat{Z}) \in
\widehat{G}_{\mathcal A}$~:
\begin{equation}
\label{bch}
    \widehat{\varphi}=\exp\big(\mathrm{R}(\chi(\widehat{Z}))\big)
                      \exp\big(\tilde{\mathrm{R}}(\chi(\widehat{Z}))\big).
\end{equation}
The matrices $\widehat{\varphi}_-:=\exp\big(-\mathrm{R}(\chi(\widehat{Z}))\big)$
and $\widehat{\varphi}^{-1}_+:=\exp\big(-\tilde{\mathrm{R}}(\chi(\widehat{Z}))\big)$
are in $\widehat{G}^{-}_{\mathcal A}$ and $\widehat{G}^{+}_{\mathcal A}$,
respectively, and solve Bogoliubov's matrix formulae in
(\ref{matrixAtkinson}). }}\end{remark}

\subsection*{Acknowledgments}

We would like to thank the organizers of the CIRM 2006 Workshop ''Renormalization and Galois Theory`` for giving us the opportunity to present parts of our recent research results. The stimulating atmosphere we found at this memorable workshop affected us considerably. Many of the newer results presented here were achieved jointly with J.~M.~Gracia-Bond\'ia, and F.~Patras and a warm thanks goes to both for constant and fruitful ongoing collaborations.

\end{document}